\documentclass[final]{siamltex}

\usepackage{color,amsmath,amssymb,mathrsfs}
\usepackage{color}
\usepackage{cite}
\usepackage{showkeys}
\usepackage{graphicx}
\usepackage{algorithmic, algorithm}
\usepackage{pgf,tikz}
\usetikzlibrary{matrix,arrows,decorations.pathreplacing,decorations.pathmorphing}
\usetikzlibrary{plotmarks,arrows,positioning}
\usepackage{pgfplots}

\usepackage[plainpages=false,colorlinks=true,citecolor=blue,
  filecolor=blue,linkcolor=blue,urlcolor=blue]{hyperref}

\definecolor{green}{rgb}{0,0.65,0}

\newcommand{\dbeta}{\ensuremath{\vec{\beta}}}
\newcommand{\uu}{\ensuremath{\vec{u}}}
\newcommand{\vv}{\ensuremath{\vec{v}}}
\newcommand{\ff}{\ensuremath{\vec{f}}}

\newcommand{\edot}{\dot{\ten{\varepsilon}}}
\newcommand{\secinve}{\edot_\mathrm{I\!\!\:I}}

\renewcommand{\matrix}[1] {\ensuremath{\boldsymbol{#1}}}
\renewcommand{\vec}[1] {\ensuremath{\boldsymbol{#1}}}
\renewcommand{\H}{\ensuremath{\matrix{H}}}
\newcommand{\M}{{\ensuremath{\matrix{M}}}}

\newcommand{\K}{{\ensuremath{\matrix{K}}}}
\newcommand{\I}{{\ensuremath{\matrix{I}}}}
\newcommand{\A}{{\ensuremath{\matrix{A}}}}
\newcommand{\B}{{\ensuremath{\matrix{B}}}}
\newcommand{\V}{{\ensuremath{\matrix{V}}}}

\renewcommand{\L}{{\ensuremath{\matrix{L}}}}
\newcommand{\F}{{\ensuremath{\matrix{F}}}}

\newcommand{\HM}{\matrix{H}_{\text{misfit}}}

\newcommand{\gbf}[1]{\ensuremath{\boldsymbol{#1}}}
\newcommand{\obs}{\vec{d}}
\newcommand{\ipar}{m}
\newcommand{\dpar}{\vec{m}}

\newcommand{\xx}{\ensuremath{\boldsymbol x}}

\newcommand{\x}{\xx}
\newcommand{\y}{\ensuremath{\boldsymbol y}}

\newcommand{\eqnlab}[1]{\label{eq:#1}}
\newcommand{\eqnref}[1]{\eqref{eq:#1}}

\definecolor{pacificorange}{cmyk}{.15,.45,1,0} %
\definecolor{pacificgray}{cmyk}{0,.15,.35,.60}
\definecolor{pacificlgray}{cmyk}{0,0,.2,.4}
\definecolor{pacificcream}{cmyk}{.05,.05,.15,0}
\definecolor{deepyellow}{cmyk}{0,.17,.80,0}
\definecolor{lightblue}{cmyk}{.49,.01,0,0}
\definecolor{lightbrown}{cmyk}{.09,.15,.34,0}
\definecolor{deepviolet}{cmyk}{.79,1,0,.15}
\definecolor{deeporange}{cmyk}{0,.59,1,.18}
\definecolor{dustyred}{cmyk}{0,.7,.45,.4}
\definecolor{grassgreen}{RGB}{92,135,39}
\definecolor{pacificblue}{RGB}{59,110,143}
\definecolor{pacificgreen}{cmyk}{.15,0,.45,.30}
\definecolor{deepblue}{cmyk}{1,.57,0,.2}
\definecolor{turquoise}{cmyk}{.43,0,.24,0}
\definecolor{darkgreen}{rgb}{0.25,0.65,0.10}

\newcommand{\bs}[1]{\boldsymbol{#1}}
\newcommand{\ten}[1] {\ensuremath{\boldsymbol{#1}}}

\newcommand{\mc}[1]{\mathcal{#1}}
\newcommand{\nor}[1]{\left\| #1 \right\|}

\newcommand{\yobs}{\obs^{\text{obs}}}
\newcommand{\R}{\mathbb{R}}
\newcommand{\half} {\ensuremath{\frac{1}{2}}}
\newcommand{\D}{\Omega}
\newcommand{\Acal}{\mc{A}}

\newcommand{\X}{X}

\newcommand{\C}{\mc{C}}

\newcommand{\DD}[2] {\ensuremath{\frac{d {#1}}{d {#2}}}}

\newcommand{\LRp}[1]{\left( #1 \right)}

\newcommand{\LRc}[1]{\left\{ #1 \right\}}
\newcommand{\GM}[2]{\mc{N}\left( #1, #2 \right)}

\newcommand{\map}{\params_{\mbox{\tiny{MAP}}}}
\newcommand{\Hmap}{\H_{\mbox{\tiny{MAP}}}}
\newcommand{\SN}{\mbox{\tiny{SN}}}
\newcommand{\SNMAP}{\mbox{\tiny{SNMAP}}}
\newcommand{\IS}{\mbox{\tiny{ISMAP}}}

\newcommand{\proposal}{\bs y}
\newcommand{\params}{\bs m}

\newcommand{\like}{ \pi_{\rm{like}} }
\newcommand{\post}{ \pi_{\rm{post}} }
\newcommand{\ncov} {\bs \Gamma_{\rm{noise}} }

\newcommand{\prcov} {\bs \Gamma_{\rm{prior}}    }

\newcommand{\Lpr}     {\bs L}

\newcommand{\cost}{J(\params)}

\newcommand{\Hpost}  { \bs H }

\newcommand{\Vr} {\bs V_r}
\newcommand{\Dr} {\bs D_r}
\newcommand{\Ir} {\bs I_r}

\newcommand{\mip}[2]{\left\langle{#1}, {#2}\right\rangle_{\M}}
\newcommand{\ip}[2]{\left\langle{#1}, {#2}\right\rangle}

\makeatletter
\def\addcontentsline#1#2#3{%
  \addtocontents{#1}{\protect\contentsline{#2}{#3}{\thepage}}}
\long\def\addtocontents#1#2{%
  \protected@write\@auxout
    {\let\label\@gobble \let\index\@gobble \let\glossary\@gobble}%
    {\string\@writefile{#1}{#2}}}

\newcommand\tableofcontents{%
    \section*{%
        \@mkboth{\noindent%
           \MakeUppercase\contentsname}{\MakeUppercase\contentsname}}%
    \@starttoc{toc}%
    }
\newcommand*\l@section[2]{%
  \ifnum \c@tocdepth >\z@
    \addpenalty\@secpenalty
    \addvspace{1.0em \@plus\p@}%
    \setlength\@tempdima{1.5em}%
    \begingroup
      \parindent \z@ \rightskip \@pnumwidth
      \parfillskip -\@pnumwidth
      \leavevmode \bfseries
      \advance\leftskip\@tempdima
      \hskip -\leftskip
      #1\nobreak\hfil \nobreak\hb@xt@\@pnumwidth{\hss #2}\par
    \endgroup
  \fi}
\newcommand*\l@subsection{\@dottedtocline{2}{1.5em}{2.3em}}
\newcommand*\l@subsubsection{\@dottedtocline{3}{3.8em}{3.2em}}
\newcommand*\l@paragraph{\@dottedtocline{4}{7.0em}{4.1em}}
\newcommand*\l@subparagraph{\@dottedtocline{5}{10em}{5em}}
\makeatother

\newcommand{\adjMacroMM}[1]{{#1}^*}
\newcommand{\adjMacroME}[1]{{#1}^\natural}
\newcommand{\adjMacroEM}[1]{{#1}^\diamond}
\newcommand{\adjMacroMMINV}[1]{{#1}^{-*}}

\newcommand{\Badj}{\adjMacroMM{\B}}

\newcommand{\Fadj}{\adjMacroME{\F}}

\newcommand{\Vadj}{\adjMacroEM{\V}}

\newcommand{\drawvarianceh}[2]{
  \begin{tikzpicture}[scale=#1]
    \begin{axis}[hide y axis, axis x line=bottom]
        \addplot[smooth, color=red, mark=none, line width=2pt]
        table[x=x,y=z]  {#2};
        \addplot[smooth, color=blue, mark=none, line width=2pt]
        table[x=x,y=y]  {#2};
      \end{axis}
    \end{tikzpicture}
}

\newcommand{\Width}{12.3}
\newcommand{\Height}{2.2}
\newcommand{\drawlegend}[3]{
  \begin{tikzpicture}[scale=#1]
    \draw [] rectangle (\Width,\Height);
    \draw[-,color=blue, line width=2pt] (0.05*#2,3*#3) -- (0.3*#2, 3*#3);
    \draw[-,color=red,  line width=2pt] (0.05*#2,6*#3) -- (0.3*#2, 6*#3);
    \node [color=black] at (1.0*#2,6*#3) {\small Estimated posterior pdf};
    \node [color=black] at (1.01*#2,3*#3) {\small Gaussian approximation};
    \end{tikzpicture}
}

\def \addressices{Institute for Computational Engineering \& Sciences, The
  University of Texas at Austin, Austin, TX, USA}

\def \addressgeo{Department of Geological Sciences, The University of
  Texas at Austin, Austin, TX, USA}

\def \addressmech{Department of Mechanical Engineering, The
  University of Texas at Austin, Austin, TX, USA}

\begin{document}

\author{Noemi Petra\footnotemark[2] \and James Martin\footnotemark[2]
  \and Georg Stadler\footnotemark[2] \and Omar
  Ghattas\footnotemark[2] \footnotemark[3] \footnotemark[4]}

\renewcommand{\thefootnote}{\fnsymbol{footnote}}
\footnotetext[2]{\addressices}
\footnotetext[3]{\addressmech}
\footnotetext[4]{\addressgeo}
\renewcommand{\thefootnote}{\arabic{footnote}}

\title{A computational framework for infinite-dimensional Bayesian
  inverse problems \\Part II: Stochastic Newton MCMC with application
  to ice sheet flow inverse problems \thanks{Support for this work
    was provided by the U.S.~National Science Foundation (NSF)
    under grant ARC-0941678, and by the U.S.~Department of
    Energy Office of Science, Advanced Scientific Computing Research
    and Biological and Environmental Research programs under grants DE-SC0009286, DE-11018096,
    DE-SC0006656, and DE-SC0002710.}}
\maketitle

\begin{abstract}
We address the numerical solution of infinite-dimensional inverse
problems in the framework of Bayesian inference. In the Part I
\cite{Bui-ThanhGhattasMartinEtAl13} companion to this paper, we
considered the linearized infinite-dimensional inverse problem. Here
in Part II, we relax the linearization assumption and consider the
fully nonlinear infinite-dimensional inverse problem using a Markov
chain Monte Carlo (MCMC) sampling method. To address the challenges of
sampling high-dimensional probability density functions (pdfs) arising
upon discretization of Bayesian inverse problems governed by PDEs,
we build on the stochastic Newton MCMC method. This method exploits
problem structure by taking as a proposal density a local Gaussian
approximation of the posterior pdf, whose covariance operator is given
by the inverse of the local Hessian of the negative log posterior
pdf. The construction of the covariance is made tractable by invoking
a low-rank approximation of the data misfit component of the
Hessian. Here we introduce an approximation of the stochastic Newton
proposal in which we compute the low-rank-based Hessian at just the
MAP point, and then reuse this Hessian at each MCMC step. We compare
the performance of the proposed method to the original stochastic
Newton MCMC method and to an independence sampler. The comparison of
the three methods is conducted on a synthetic ice sheet inverse
problem. For this problem, the stochastic Newton MCMC method with a
MAP-based Hessian converges at least as rapidly as the original
stochastic Newton MCMC method, but is far cheaper since it avoids
recomputing the Hessian at each step. On the other hand, it is more
expensive per sample than the independence sampler; however, its
convergence is significantly more rapid, and thus overall it is much
cheaper. Finally, we present extensive analysis and interpretation of
the posterior distribution, and classify directions in parameter space
based on the extent to which they are informed by the prior or the
observations.
\end{abstract}



\pagestyle{myheadings}
\thispagestyle{plain}
\markboth{N.\ Petra, J.\ Martin, G.\ Stadler, O.\ Ghattas}
{STOCHASTIC NEWTON MCMC FOR BAYESIAN INVERSE PROBLEMS}

\section{Introduction and background}
\label{sec:introduction}

We consider the problem of estimating the uncertainty in the solution
of infinite-dimensional inverse problems within the framework of
Bayesian inference \cite{Stuart10, Tarantola05,
  KaipioSomersalo05}. Namely, given observational data and their
uncertainties, a (possibly stochastic) forward model that maps model
parameters to observations, and a prior probability distribution on
model parameters that encodes any prior knowledge or assumptions about
the parameters, find the {\em posterior probability distribution} of
the parameters conditioned on the observational data. This probability
density function (pdf) is defined as the Bayesian solution of the
inverse problem.  The posterior distribution assigns to any candidate
set of parameter fields our belief (expressed as a probability) that a
member of this candidate set is the ``true'' parameter field that gave
rise to the observed data.

The standard approach to explore the posterior distribution is based
on sampling using a Markov chain Monte Carlo (MCMC) method.  However,
the use of conventional MCMC methods becomes intractable for
large-scale inverse problems, which arise upon discretization of
infinite-dimensional inverse problems.  This is due to the twin
difficulties of high dimensionality of the uncertain parameters and
computationally expensive forward models.

A number of methods have
emerged to address
Bayesian inverse problems governed by PDEs (we give a representative
recent reference in each case, which can be consulted for additional
references to historical work; further references can be found in the
recent survey \cite{FoxHaarioChristen12}): replacing the forward
problem with a reduced order model in both parameter and state space
\cite{LiebermanWillcoxGhattas10}; approximating the
parameter-to-observable map \cite{HigdonGattikerWilliamsEtAl08}
or the posterior \cite{Bui-ThanhGhattasHigdon12} with a Gaussian
process response surface; employing a polynomial chaos approximation
of the forward problem \cite{MarzoukNajm09}; using a two-stage
``delayed acceptance'' MCMC method in which the first stage employs an
approximate forward model \cite{CuiFoxOSullivan12};
employing gradient information (of the
negative log posterior) to accelerate sampling, as in
Langevin methods 
\cite{DostertEfendievHouEtAl06,StuartVossWiberg04,RobertsTweedie96}
and
their preconditioned variants 
\cite{BeskosRobertsStuart09}; exploiting Riemannian geometry of
parameter space to accelerate sampling \cite{GirolamiCalderhead11};
and creating an MCMC proposal that uses local gradient and
low-rank Hessian information of the negative log posterior to
construct a local Gaussian approximation
\cite{MartinWilcoxBursteddeEtAl12}.

Here we focus on the last of these methods,
the so-called {\em stochastic Newton MCMC method}.
This method employs local Hessian-based Gaussian proposals 
that exploit the structure of the underlying
posterior to guide the sampler to regions with higher acceptance
probability.  In particular, such proposals capture the highly
stretched contours of the posterior that are typical for ill-posed
inverse problems, in which the data inform the model parameters very
well in some directions in parameter space, and poorly in others.
One of the challenges in employing the Hessian is that
its explicit construction  entails solution of as many
forward problems as there are parameters, which is 
out of the question for large-scale forward problems. These
difficulties are addressed by introducing low-rank approximations of
the Hessian, motivated by the compact nature of the Hessian operator
for many inverse problems~\cite{Bui-ThanhBursteddeGhattasEtAl12,
  FlathWilcoxAkcelikEtAl11, MartinWilcoxBursteddeEtAl12,
  Bui-ThanhGhattasMartinEtAl13, Bui-ThanhGhattas12,
  Bui-ThanhGhattas12a, Bui-ThanhGhattas12f}. This delivers accurate
approximation of the Hessian at a cost that is independent of the
parameter dimension (when the parameter represents a discretized
field), leading to  solution of Bayesian inverse problems with
non-trivial dimensions \cite{MartinWilcoxBursteddeEtAl12}. 
Other work employing Hessian-based proposals includes
the taylored chain approach
\cite{GewekeTanizaki03}, and, specifically, in the context of nonlinear
filtering \cite{GewekeTanizaki99}; the Hessian-based
Metropolis-Hastings (HMH) algorithm with a learning rate to influence step
size \cite{QiMinka02}; a position-specific preconditioned Metropolis
adjusted Langevin algorithm
(PSP-MALA) implemented with a block Metropolis-Hastings algorithm \cite{Herbst10};
function-space MCMC proposals for which the prior is invariant,
and thus insensitive to mesh refinement \cite{Law13};
and, finally, the Random Maximized Likelihood (RML) algorithm which
generates samples as the solutions of related deterministic inverse problems
\cite{OliverReynoldsLiu08}.

Despite the low-rank approximation, stochastic Newton MCMC (and any
method that uses local Hessian information) is computationally
expensive for large-scale problems, since at every proposed sample
point the gradient and a low-rank approximation of the Hessian are
computed, which requires multiple forward and adjoint PDE solves
having the same linear operator (or its adjoint). When these PDE
solves are done iteratively, there is little opportunity to exploit
the fact that the linear operators are the same, beyond amortizing the
cost of preconditioner construction over the solves. 

To alleviate this computational cost, here we propose a modified
stochastic Newton MCMC that uses proposals based on local gradient
information as well as on Hessian information computed initially at
the maximum a posteriori (MAP) point and then reused at every sample
point.  We call this the {\it stochastic Newton MCMC method with
  MAP-based Hessian}.  We compare this proposed method with the
original stochastic Newton MCMC method (with dynamically-computed
Hessian) as well as with
an independence sampler that uses a Gaussian proposal centered at the
MAP point, using the Hessian computed at the MAP as the covariance
\cite{OliverReynoldsLiu08}. This independence sampler is
computationally attractive since (like the proposed stochastic Newton MCMC 
method with MAP-based Hessian), the Hessian is computed just once, but
(unlike the new method), the gradient is used only to determine the
MAP. Because the proposed stochastic Newton MCMC method with MAP-based
Hessian uses local (gradient) information, we expect it will
outperform the independence sampler; because it freezes the Hessian at
the MAP point, it will be significantly cheaper per sample than the
original stochastic Newton MCMC method.

The stochastic Newton MCMC method with MAP-based Hessian can be
derived as a particular variant of a preconditioned
Metropolis-adjusted Langevin algorithm using preconditioning based on
the Hessian at the MAP point. Note that all of the above methods
attempt to exploit problem structure---in particular the local
curvature of the posterior---by making use of Hessian information to
one degree or another.  Note also that all three of these
Hessian-based methods reduce to the same method when the target
inverse problem is linear and the prior and noise pdfs are Gaussian
(in which case the posterior is also Gaussian). For non-Gaussian
posteriors, however, the three methods take distinct steps.

Beyond this new, more efficient, variant of stochastic Newton MCMC, this
article extends our previous work on methods for large-scale Bayesian
inverse problems~\cite{Bui-ThanhGhattasMartinEtAl13,
  MartinWilcoxBursteddeEtAl12} in several directions.  In
\cite{Bui-ThanhGhattasMartinEtAl13}, we presented a computational
framework for linearized infinite-dimensional Bayesian inverse
problems, building on the infinite-dimensional formulation of
Stuart~\cite{Stuart10}. Here, we extend our computational framework to
nonlinear inverse problems, for which the posteriors are non-Gaussian,
requiring MCMC sampling.
To this end, we extend the finite-dimensional stochastic Newton MCMC method
presented in \cite{MartinWilcoxBursteddeEtAl12} to be consistent with
the infinite-dimensional setting. This requires care in discretizing
the prior and likelihood and establishing finite-dimensional inner
products, which arise in multiple steps of stochastic Newton.

We study the efficiency of the proposed method in the context of an
ice sheet flow Bayesian inverse problem, in which a basal
boundary condition parameter field is inferred from surface velocity
observations. Here, the parameter-to-observable map involves the
solution of a nonlinear Stokes equation describing viscous, creeping,
incompressible, non-Newtonian ice flow.
This extends recent research on ice sheet inverse problems, which
focused on deterministic inversion or the computation of the MAP
solution~\cite{PetraZhuStadlerEtAl12, GoldbergSergienko11,
  MorlighemRignotSeroussiEtAl10, PralongGudmundsson11,
  RaymondGudmundsson09, PricePayneHowatEtAl11}. We apply the full
Bayesian inference framework and study the performance of the three
Hessian-based methods described above in exploring the posterior pdf.
Convergence of the three methods is studied using various diagnostics
to assess MCMC chain convergence. We also compare with a reference
Delayed Rejection Adaptive Metropolis (DRAM)
sampler~\cite{HaarioLaineMiraveteEtAl06} that, similar to stochastic
Newton, attempts to capture the curvature of the posterior, but
without relying on gradient or Hessian information. The results reveal
that, among the Hessian-based methods, the stochastic Newton MCMC method
with MAP-based Hessian yields the fastest convergence in terms of both
the number of samples and the computational work. In comparison, DRAM
is incapable of making progress on this problem.

Finally, we study and interpret visually the solution of the Bayesian
inverse problem with respect to the information contained in the data
and in the prior and the effect they have on the posterior in high
dimensions. 
This can be challenging in high dimensions, but we demonstrate
that it can be made tractable by exploiting knowledge contained within 
the spectral structure of the Hessian of the log likelihood 
evaluated at the MAP point as well as the prior covariance. 
Because this structure is common to many Bayesian inverse problems, we
expect that these strategies for visualization will be of general
value beyond the specific application.

The remaining sections of this paper are organized as follows. We
begin by providing in Section~\ref{subsec:infbayesianform} 
an overview of the framework for
infinite-dimensional Bayesian inverse problems
following~\cite{Bui-ThanhGhattasMartinEtAl13, Stuart10}. Next,
in Section~\ref{subsec:discretebayesianform} we present a consistent
discretization of the infinite-dimensional inverse
problem. Section~\ref{sec:proposals} presents the proposed stochastic
Newton MCMC method with MAP-based Hessian, while
Section~\ref{subsec:lra} describes our low rank-based Hessian
approximation. Section~\ref{sec:app} 
introduces a Bayesian formulation of an ice sheet flow inverse
problem, and gives expressions for adjoint-based gradient and
Hessian-vector products (of the negative log posterior). In
Section~\ref{sec:uq}, we discuss 
the performance of the three sampling methods. Finally, in
Section~\ref{sec:uqinterpr} we interpret the posterior distribution by
visualizing marginals with respect to the eigenvectors of the
covariance operator. This provides insight into the ability of the
observations to infer model parameters. Section~\ref{sec:conclusions}
provides concluding remarks.

\section{Background on the infinite-dimensional Bayesian inverse
  problem, its consistent discretization, and characterization of the
  posterior}
\label{sec:prelim}

Formulating and solving the Bayesian inverse problem for an
infinite-dimensional parameter field presents difficulties. First, the
usual notion of a pdf is not defined
since there is no Lebesgue measure in infinite dimensions. Second, the
prior measure must be chosen appropriately to lead to a well-posed
inverse problem and facilitate computation of the posterior.  Third,
the choice of discretization must be consistent with the
infinite-dimensional structure of the problem. Finally, exploring the
posterior that arises upon discretization via an MCMC method is
typically prohibitive due to the resulting high dimensionality of the
parameter space.

In this section we formulate the Bayesian inverse
problem in infinite dimensions (Section~\ref{subsec:infbayesianform}) in the
framework of \cite{Stuart10}, which uses the  Radon-Nikodym
derivative and an appropriately chosen Gaussian prior that employs as
covariance operator the inverse of an elliptic differential operator. 
In Section~\ref{subsec:discretebayesianform}, we describe the
discretization of this infinite-dimensional inverse problem in a way
that is consistent with the underlying infinite-dimensional function
spaces. This leads to non-standard definitions of operator
adjoints. 
When the posterior is nearly Gaussian, its mean and covariance can be
approximated by the MAP point and the inverse of the Hessian evaluated
at the MAP. 
Inversion of the Hessian is intractable in high dimensions; 
Section~\ref{subsec:lra} presents a low-rank approximation of the Hessian
of the data misfit in order to make these Hessian computations tractable. 
When the posterior is not approximately Gaussian, the method of choice
is often to sample it with an MCMC method and then compute sample
statistics; Section~\ref{subsec:mcmc} gives an overview of MCMC methods for
sampling posteriors.
 
\subsection{Bayesian formulation of infinite-dimensional inverse problems}
\label{subsec:infbayesianform}

In an inverse problem, we seek to infer the unknown (or uncertain)
input parameters to a mathematical model from observations of the
outputs of the model. For ill-posed inverse problems, the uncertain
parameter $\ipar \in \mathcal H$ is often a heterogeneous field over a
domain $\Omega$, and $\mathcal H$ is typically a subset of
$L^2(\Omega)$.  The mathematical model is characterized by the
parameter-to-observable map $\ff : \mathcal H \rightarrow \mathbb
R^q$, which predicts observables $\y \in \mathbb R^q$ corresponding to
a given parameter $\ipar$. Note that this map involves solution of the
forward problem, typically a system of PDEs, followed by an
application of an observation operator.
We assume here that the observables $\y$ are
finite-dimensional. 
Given observation data $\yobs \in \mathbb R^q$, the solution to the
inverse problem seeks parameters $\ipar$ such that
\begin{equation*}
  \ff\LRp{\ipar} \approx \yobs \label{eq:invpb}
\end{equation*}
in a sense made precise by the Bayesian formulation described next.

The Bayesian formulation poses the inverse problem as a problem of
statistical inference over parameter space. The solution of the
resulting \emph{Bayesian inverse problem} is a probability
distribution that represents our belief about the correct value of the
parameter.
Solving the inverse problem using Bayes' approach requires
specification of a \emph{prior model}, which describes our
beliefs about the parameter before any data are considered, and a
\emph{likelihood model}, which quantifies the relative
probability that a candidate parameter $\ipar$ could have
produced the observed data $\yobs$.

Here we present a summary of the discussion in
\cite{Bui-ThanhGhattasMartinEtAl13}. 
The prior is taken to be the Gaussian measure $\mu_0 =
\GM{\ipar_0}{\C_0}$ on $L^2(\Omega)$, where $\ipar_0 \in \mathcal H$,
and $\C_0$ is an appropriate covariance operator $\C_0$; in
particular, $\C_0$ must be symmetric, positive, and of trace-class
\cite{Stuart10}.  We choose the covariance operator to be the inverse
of an elliptic differential operator $\mc{A}$ that is of sufficiently
high order to guarantee a well-posed Bayesian inverse problem
\cite{Stuart10}.
We choose $\mc{A}$ to be second order differential operator\footnote{
  The necessary order of $\mc{A}$ to lead to a valid covariance operator depends on the spatial dimension of
  the domain $\Omega$ \cite{Stuart10}.  In the example considered in Section~\ref{sec:app}, the inversion parameter
  is a one-dimensional field, and a second order differential operator
  is sufficient to guarantee that $\C_0$ is a valid covariance
  operator.
  While there is no distinction in one dimension
  between ordinary and partial derivatives, we choose to express
  $\mc{A}$ in the language of PDEs for notational consistency with the
  development in \cite{Bui-ThanhGhattasMartinEtAl13}.}
expressed in weak form:
for $s\in
L^2(\Omega)$, the solution $\ipar=\mc A^{-1}s$ satisfies
  \begin{equation}
    \eqnlab{Wspace}
    \int_\Omega [a \nabla \ipar \cdot \nabla p + b \ipar p] \,d\xx = \int_\Omega
    sp\,d\xx \quad \text{ for all } p \in H^1(\Omega),
  \end{equation}
  with $a, b > 0$. These coefficients control the correlation length
  and the variance in the covariance operator $\mc A^{-1}$. Choosing
  for spatially dependent coefficients $a$ and $b$ or a tensor
  coefficient $a$ allows the incorporation of further problem specific
  knowledge, such as spatially varying or anisotropic correlations, in
  the covariance operator $\mc
  A^{-1}$~\cite{Bui-ThanhGhattasMartinEtAl13}.

For the likelihood model, we assume  that observational
uncertainty (i.e., uncertainty in $\yobs$ related to measurement
error) and model uncertainty (i.e., uncertainty in $\ff(\ipar)$ due to
inadequacy of the forward model) are each centered, additive, and
Gaussian. We combine these into a single \emph{noise model},
\begin{equation*}
\ff(m) = \y + \vec \eta , \qquad \text{with} \qquad \vec \eta \sim
         \GM{\vec0}{\bs{\Gamma}_\text{noise}}, 
\end{equation*}
where $\vec \eta \in \mathbb{R}^q$ is a random variable representing
noise, and $\bs{\Gamma}_\text{noise}  \in \mathbb{R}^{q \times q}$ is
the noise covariance matrix. 
We can then express the pdf for the likelihood model explicitly as
\begin{equation}
\eqnlab{likelihoodInf}
\pi_{\text{like}}(\yobs | \ipar)
  \propto \exp \left[-\frac12 ( \ff\LRp{\ipar} -
  \yobs)^T \bs{\Gamma}_\text{noise}^{-1} (\ff\LRp{\ipar} - \yobs) \right].
\end{equation}

Bayes' theorem in infinite dimensions is expressed using the 
Radon-Nikodym derivative $\DD{\mu^y}{\mu_0}$ of the
posterior measure $\mu^y$ with respect to the prior measure $\mu_0$,
\begin{equation}\label{eq:BayesianSolution}
\DD{\mu^y}{\mu_0} = \frac{1}{Z} \pi_{\text{like}}(\yobs | \ipar),
\end{equation}
where $Z = \int_{\X}\pi_{\text{like}}(\yobs|\ipar) \, d\mu_0$ is a
normalization constant. For technical conditions under which the
posterior measure is well defined, and a discussion of the Bayes rule
for probability measures on function spaces, we refer the reader
to~\cite{CotterDashtiRobinsonEtAl09,CotterDashtiStuart10,Stuart10}.

\subsection{Discretization of the Bayesian inverse problem}
\label{subsec:discretebayesianform}

In this section, we present a brief discussion of the
finite-dimensional approximations of the prior and the posterior
distributions; a lengthier discussion can be found in
\cite{Bui-ThanhGhattasMartinEtAl13}.
We start with a finite-dimensional subspace $V_h$ of $L^2(\Omega)$
originating from a finite element discretization with continuous
Lagrange basis functions
$\LRc{\phi_j}_{j=1}^n$~\cite{BeckerCareyOden81,StrangFix88}. The
approximation of the inversion parameter function $\ipar \in
L^2(\Omega)$ is then $\ipar_h = \sum_{j=1}^nm_j\phi_j \in V_h$, where
the vector of the $n$ inversion parameters is $\bs m =
\LRp{m_1,\hdots,m_n}^T\in \R^n$.

Since we postulate the prior Gaussian measure on $L^2\LRp{\D}$, the
finite-dimensional space $V_h$ inherits the $L^2$-inner product. Thus,
inner products between nodal coefficient vectors must be weighted by a
mass matrix $\M\in \R^{n\times n}$ to approximate the infinite-dimensional $L^2$-inner
product. 
This $M$-weighted inner product is denoted by $\mip{\cdot\,}{\cdot}$,
where $\mip{\vec{y}}{\vec{z}} = \vec{y}^T \M \vec{z}$ and $\M$ is the
(symmetric positive definite) mass matrix
\[
M_{ij} = \int_\Omega \phi_i(\x) \phi_j(\x) \, d\x ~, \quad i,j =
1,\hdots,n.
\]
To distinguish $\R^n$ equipped with the $M$-weighted inner product
with the usual Euclidean space $\R^n$, we denote it by $\R^n_\M$.

When using the $M$-weighted inner product, there is a critical distinction that
must be made between the matrix adjoint and the matrix transpose. For
an operator $\B: \R^n_{\M} \rightarrow \R^n_{\M}$, 
we denote the matrix transpose by $\B^T$ with entries $(B^T)_{ij} =
B_{ji}$.  In contrast, the $M$-weighted inner product adjoint $\B^*$ satisfies, for $\vec y, \vec z\in \R^n$,
\[
\mip{\B\vec y}{\vec z} = \mip{\vec y}{\B^* \vec z},
\]
which implies that $\B^*$ is given by
\begin{align}
  \label{eq:MM_adj}
  \Badj &= \M^{-1} \B^T \M.
\end{align}
In the following, we also need the adjoint
$\Vadj$ of $\V : \R^r \rightarrow \R^n_{\M}$
(for some $r$), where $\R^r$ is endowed with the
Euclidean inner product.
In this case, we have
\begin{align}
\label{eq:EM_adj}
\Vadj &= \V^T \M,
\end{align}
since $\mip{\V \vec y}{\vec z} = \ip{\vec y}{\Vadj \vec z}$.
With these definitions, the matrix representation of the elliptic PDE operator
$\Acal$ defined by \eqnref{Wspace} is given by $\A = \M^{-1}
\K\in \R^{n\times n}$~\cite{Bui-ThanhGhattasMartinEtAl13}, where $\K\in \R^{n\times n}$ is the stiffness
matrix
\[
K_{ij} = \int_\Omega [a \nabla\phi_i(\x) \cdot \nabla\phi_j(\x) + b
\phi_i(\x)\phi_j(\x)] \, d\x, \quad i,j \in \LRc{1,\hdots,n}.
\]
Then, the finite-dimensional approximation $\mu_0^h$ of the prior Gaussian
measure $\mu_0$ is the more familiar multivariate Gaussian 
with density
\begin{equation}
  \label{eq:prior_pdf}
  \pi_{\text{prior}}(\dpar) \propto \exp\left[
    -\frac{1}{2}\mip{\dpar-{\dpar}_0}{\A(\dpar-{\dpar}_0)}\right],
\end{equation}
where $\dpar_0\in \R^n$ is the discretization of the prior mean $\ipar_0$.
The finite-dimensional Bayes' formula, i.e.,
\begin{align}
\label{eq:new_finite_bayes}
\pi_{\text{post}}(\dpar):=\pi_{\text{post}}(\dpar | \yobs) \propto
\pi_{\text{prior}}(\dpar) \pi_{\text{like}}(\yobs | \dpar),
\end{align}
where $\pi_{\text{post}}(\dpar | \yobs)$ is the density of the
finite-dimensional approximation $\mu^{y,h}$ of the posterior measure
$\mu^y$, and $\pi_{\text{like}}$ is the likelihood
\eqnref{likelihoodInf},
gives the finite-dimensional posterior
density explicitly as
\begin{equation}
\label{eq:explicit_finite_posterior}
\pi_{\text{post}}(\dpar)
\propto
\exp\left[
-\frac 12 \left\| \ff(\dpar) - \yobs\right\|^2_{\matrix{\Gamma}^{-1}_{\text{noise}}}
\!-\frac{1}{2}\mip{\dpar-{\dpar}_0}{\bs{\Gamma}_\text{prior}^{-1}(\dpar-{\dpar}_0)}
\right],
\end{equation}
where $\bs{\Gamma}_\text{prior} = \A^{-1}$. Note that in
\eqref{eq:explicit_finite_posterior} and the remainder of this paper
we denote by $\ff(\dpar)$ the parameter-to-observable map evaluated at
the finite element function corresponding to the parameter vector $\dpar$.
The Bayesian solution of the inverse problem is then given by
\eqref{eq:explicit_finite_posterior}. Unfortunately, for inverse
problems governed by expensive forward models and for high-dimensional
parameter spaces, exploring the posterior density
$\pi_{\text{post}}(\dpar)$ is extremely challenging, since evaluation
of this density at any point in parameter space requires the solution
of the forward model $\ff(\dpar)$ for the given $\dpar$, and a very
large number of such evaluations will be required in high dimensions.
Methods for exploring $\pi_{\text{post}}(\dpar)$ that do not exploit
its structure are thus impractical.

We observe that the negative log posterior density is analogous to the
least squares functional 
that is minimized in the solution of a deterministic inverse problem. 
That is,
\begin{equation}
  \label{eq:posterior_V}
- \log \post(\dpar)  = J(\params) + \mbox{ const.}
\end{equation}
where
\begin{equation}\label{eq:cost}
  \cost := \tfrac{1}{2} \| \ff(\dpar) -
  \yobs \|^2_{\ncov^{-1}} +
  \frac{1}{2}\mip{\dpar-{\dpar}_0}{\bs{\Gamma}_\text{prior}^{-1}(\dpar-{\dpar}_0)}.
\end{equation}
In the context of deterministic inversion, the first term in
\eqref{eq:cost} is the data misfit term, weighted by $\ncov^{-1}$, and
the second term plays the role of Tikhonov regularization, which is
chosen to make the inverse problem well-posed. This connection between
the negative log posterior and the deterministic inverse problem cost
function in \eqref{eq:cost} is often exploited to find an
approximation of the mean of the posterior pdf by finding the point
that maximizes the posterior $\pi_{\text{post}}(\dpar)$, or
equivalently minimizes the cost function $ J(\params)$. This so-called
maximum a posterior (MAP) point is equal to the mean when the
parameter-to-observable map $\ff(\dpar)$ is linear in the parameters
$\dpar$ and the noise and prior models are Gaussian. When the
Gaussian-linear conditions are not satisfied, obviously the MAP point
only approximates the mean, the quality of this approximation
depending on the degree of nonlinearity.  Moreover, under these
Gaussian-linear conditions, the posterior $\pi_{\text{post}}(\dpar)$
is Gaussian with mean given by the MAP point, and covariance given by
the inverse of the Hessian matrix of the cost function $ J(\params)$ 
\cite{Tarantola05, Bui-ThanhGhattasMartinEtAl13}.

\subsection{Exploring the posterior}
\label{subsec:mcmc}

As implied above, when the parameter-to-ob\-ser\-va\-ble map is nonlinear,
the posterior  $\pi_{\text{post}}(\dpar)$ generally is non-Gaussian,
and cannot be represented by its mean and covariance. Thus it must be
characterized by other means. This can be extremely challenging for
PDE-based inverse problems, since evaluating the posterior
\eqref{eq:explicit_finite_posterior} 
at any point in parameter space involves solving the forward PDEs, and
many such evaluations are anticipated for the high-dimensional
parameter spaces that stem from discretization of infinite-dimensional
inverse problems. 

The method of choice for exploring the posterior pdf is the
Metropolis-Hastings (M-H) MCMC
method~\cite{MetropolisRosenbluthRosenbluthEtAl53, Hastings70,
  RobertCasella04, Tierney94}, which employs a given proposal
probability density $q(\params_k,\proposal)$ at each sample point
$\params_k$ in parameter space to generate a proposed sample point
$\proposal\in \R^n$. Once generated, the M-H criterion chooses to either
accept or reject the proposed sample point, and repeats from the new
point, thereby generating a chain of samples $\{\params_k\}_{k=1,\cdots}$ from the
posterior density $\post(\params)$. Algorithm~\ref{algorithm:M-H}
presents pseudo-code for the M-H MCMC method.
\begin{algorithm}[tbp]
  \caption{Metropolis-Hastings MCMC algorithm to sample the pdf $\pi$}
	\begin{algorithmic}
	\STATE Choose initial parameters $\params_0$
	\STATE Compute $\pi(\params_0)$
	\FOR{$k = 0,\ldots,N-1$}
		\STATE Draw sample $\proposal$ from the proposal density
                $q(\params_k, \cdot \, )$
        	\STATE Compute $\pi(\proposal)$
		\STATE Compute $\alpha_k(\proposal) = \min \left\{ 1,
                               \frac{\pi(\proposal) q(\proposal,\params_k)}
			            {\pi(\params_k) q(\params_k,\proposal)}
                               \right\}$
		\STATE Draw $u \sim \mathcal U([0,1])$
		\IF{ $u < \alpha_k(\proposal)$}
			\STATE Accept: Set $\params_{k+1} = \proposal$
		\ELSE
			\STATE Reject: Set $\params_{k+1} = \params_k$
		\ENDIF
	\ENDFOR
	\end{algorithmic}
  \label{algorithm:M-H}
\end{algorithm}

Critical to the success of M-H MCMC is the choice of the proposal
density $q(\params_k,\proposal)$. Observe that if
$q(\params_k,\proposal) = \pi_{\text{post}}(\proposal)$, the M-H
algorithm would accept every sample with probability 1; however, this
defeats the purpose, because we would not know how to sample from this
choice of proposal: the whole point of appealing to MCMC is that
we cannot draw a sample directly from $\pi_{\text{post}}(\proposal)$.  
 
Instead, a common choice for the proposal is the isotropic Gaussian,
\begin{displaymath}
 q^{\mbox{\tiny{RWMH}}}(\params_k,\proposal) = 
\tfrac 1{(2\pi)^{n/2}} \exp [ -\tfrac 12 (\| \params_k - \proposal
  \|)^2 ].
\end{displaymath}
The resulting method is known as Random Walk
Metropolis-Hastings (RWMH).  This proposal density is easy to sample,
but it can lead to poor MCMC performance due to the mismatch between
the proposal and posterior densities. The challenge is to come up with
a proposal that at least locally reflects the behavior of the target
posterior density and at the same time is easy to sample. Satisfying
these two requirements becomes increasingly difficult with increasing
parameter dimension. This will be the subject of the next section.

\section{A modified stochastic Newton MCMC method}
\label{sec:proposals}

In \cite{MartinWilcoxBursteddeEtAl12}, we
introduced a so-called stochastic Newton MCMC method that featured a
Gaussian proposal constructed from the local gradient vector and local
Hessian matrix (of the negative log posterior). To make the
construction of the proposal tractable, we employed adjoint-based methods to
compute the gradient and Hessian, which amount to a pair of
forward/adjoint PDE solves for the gradient and for each column of the
Hessian. Moreover, to make the Hessian computation scalable with
respect to parameter dimension, we use matrix-free methods to construct low-rank
approximations of the data misfit component of the Hessian, which
often has a rapidly-decaying spectrum reflecting the ill-posedness of
the inverse problem \cite{MartinWilcoxBursteddeEtAl12,
  FlathWilcoxAkcelikEtAl11}. With these features, the stochastic
Newton method is able to handle inverse problems with hundreds to
thousands of parameters; its efficiency increases with decreasing
nonlinearity of the parameter-to-observable map and with decreasing
information content of the data. We denote this original form of
the stochastic Newton MCMC method (i.e., with dynamically changing
Hessian) as {\em SN}.

Unfortunately, SN becomes prohibitive for very large-scale problems,
because it requires recomputation of the Hessian at each sample
point. Despite the use of efficient adjoint-based matrix-free
Hessian-vector products to find the low-rank approximations of the
data misfit component of the Hessian, we still need $O(2r)$ linearized
forward/adjoint PDE solves to compute it, where $r$ is the effective
rank. When $r$ is large---as is the case for high-dimensional problems,
for which the observations are highly informative about the
parameters and, hence, the data misfit Hessian has a
high effective rank---we must find alternatives to computing the
Hessian at each sample point.

Here we propose a modified stochastic Newton MCMC method that employs
a (low-rank approximation-based) Hessian that is computed once and for
all at the MAP point, and reused for each proposal. This modification,
to which we refer as {\em stochastic Newton MCMC with MAP-based
  Hessian (SNMAP)}, employs a locally-computed gradient in the
Gaussian proposal, but evaluates the Hessian in that Gaussian at the
MAP point. Before describing SNMAP, we begin with a brief summary of
the proposal construction for the original stochastic Newton MCMC
method.

\subsection{Stochastic Newton MCMC with dynamically changing Hessian (SN)} 

\begin{figure}[!]
\begin{center}
  \includegraphics[width=0.45\columnwidth]{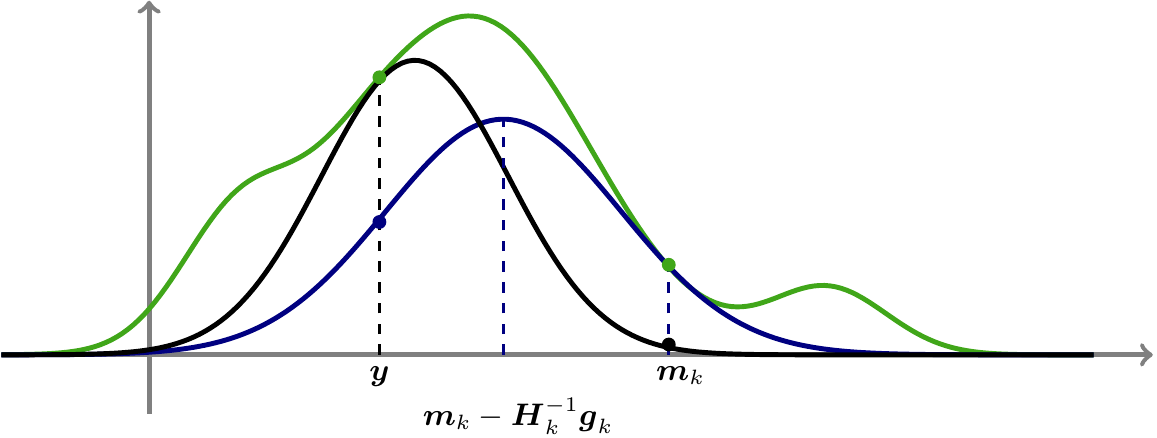}
  \includegraphics[width=0.45\columnwidth]{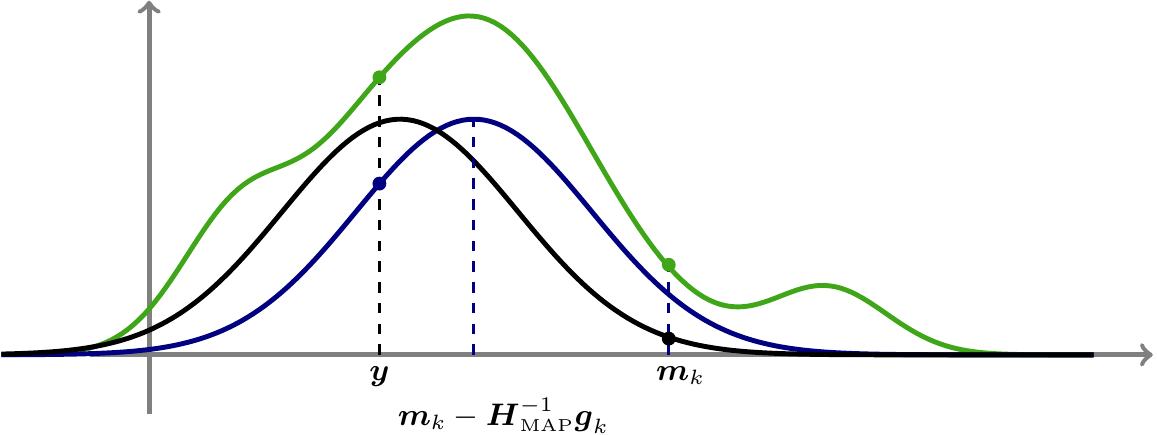}
  \includegraphics[width=0.45\columnwidth]{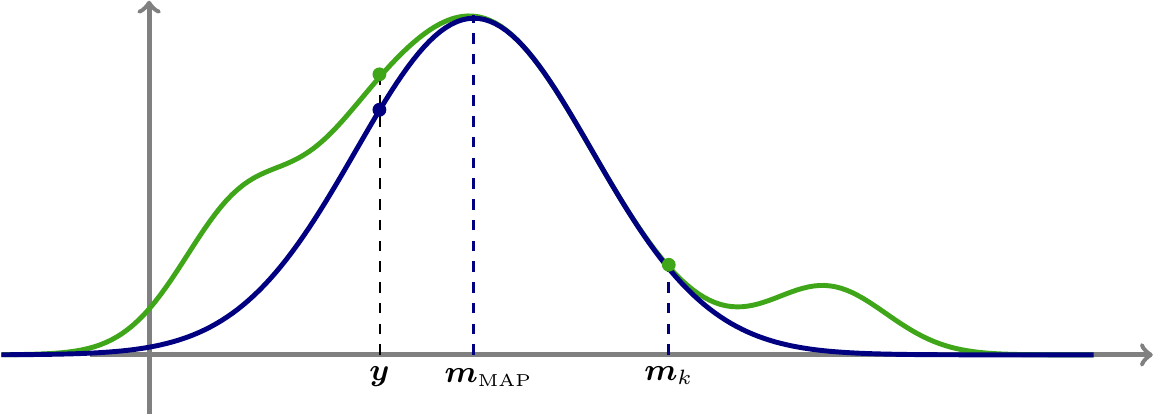}
\end{center}
\caption{Illustration of proposals for the three Hessian-based
  methods: stochastic Newton MCMC with dynamically-computed Hessian
  (top left); stochastic Newton MCMC with MAP-based Hessian (top
  right); and independence sampler with MAP-based Hessian
  (bottom). The green curve depicts the true posterior density,
  $\post(\params)$; the blue curve displays the forward proposal density,
  $q(\params_k,\proposal)$; and the black curve shows the backward
  proposal density, $q(\proposal, \params_k)$. The green, blue, and
  black dotted lines indicate the points at which the posterior, the
  backward, and the forward proposals, respectively, are evaluated.}
\label{fig:proposals}
\end{figure}

The stochastic Newton MCMC method employs a local Gaussian approximation of
the target posterior pdf. This is done by constructing, about a given
point $\params_k$, a local quadratic approximation $\tilde{J}_k(\params)$ of
the negative log posterior $J(\params)$ (given in \eqref{eq:cost}),
i.e.,
\begin{equation}
  \tilde{J}_k(\params) :=
  J(\params_k) + \mip{\bs
  g_k}{\params - \params_k} + \frac 12 \mip{\params - \params_k}{\bs
    H_k (\params - \params_k)}.
  \label{quadapprox}
\end{equation}
Here, $\bs g$ and $\bs H$ are the
gradient vector and Hessian matrix of $J(\params)$, respectively, and
$\bs g_k := \bs g(\params_k)\in \R^n$ and $\H_k := \H(\params_k)\in \R^{n\times n}$.
Rearranging terms, 
\[
\tilde{J}_k(\params)
= \frac 12 \mip{\params - \params_k + \H_k^{-1} \bs
  g_k}{\H_k (\params - \params_k + \H_k^{-1} \bs g_k)}+
\rm{const.}
\]
To obtain the proposal density $q^{\SN}$ for stochastic Newton MCMC
(with dynamically changing Hessian), we take the exponential of the negative
of $\tilde{J}_k(\params)$, and compute the scaling factor to make it a
proper pdf. This leads to
\begin{equation}
  q^{\SN}(\params_k,\proposal) = %
  \frac { \det \H_k^{1/2} }{(2\pi)^{n/2} }
  \exp
  \left(
  - \frac 12 \mip{\proposal - \params_k + \H_k^{-1} \bs g_k}{\H_k (\proposal
    - \params_k + \H_k^{-1} \bs g_k)}\right),
  \label{eq:local_q}
\end{equation}
which is a Gaussian with mean $\params_k - \H_k^{-1} \bs g_k$ and
covariance matrix $\H_k^{-1}$. Note that at a local
minimum, $\H_k$ is positive semi-definite and at an arbitrary point
$\proposal$, $\H_k$ can be indefinite.
To ensure that \eqref{eq:local_q} defines a proper pdf,
we discard negative eigenvalues of the data misfit component of $\H_k$ and, hence, replace
$\H_k$ with a modified positive definite Hessian.  We also note that
the backward proposal $q^{\SN}(\proposal, \params_k)$, needed for the
M-H acceptance probability $\alpha_k$, is computed using the Hessian
and gradient evaluated at $\proposal$.  In summary, the SN step at
each MCMC iteration draws a proposed sample $\bs y$ from the proposal
$q^{\SN}(\params_k,\proposal)$, which is then subject to the
accept/reject framework of the M-H MCMC
Algorithm~\ref{algorithm:M-H}. The SN proposal is illustrated in
Figure~\ref{fig:proposals} (top left).

\subsection{Stochastic Newton MCMC with MAP-based Hessian (SNMAP)}
\label{subsec:SNMAP}

As stated above, the original form of the stochastic Newton MCMC method
becomes prohibitive for very large-scale problems, because it requires
recomputation of the Hessian of $J(\params)$ (whose inverse is needed
to construct the Gaussian proposal) at each sample point. Therefore,
we avoid recomputing this Hessian by the following modification: we
first find the MAP point and compute the Hessian there, and then use
this MAP-based Hessian for all proposals. The gradient is still
computed at each sample point. Hence, the proposal 
$q^{\SNMAP}(\params_k,\proposal)$ for SNMAP is given by replacing the Hessian
in~\eqref{eq:local_q} with the Hessian evaluated at the MAP point,
$\Hmap$. This leads to
\begin{equation}
  q^{\SNMAP}(\params_k,\proposal) \propto
  \exp
  \left(
  - \frac 12 \mip{\proposal - \params_k + \Hmap^{-1} \bs
    g_k}{\Hmap (\proposal
    - \params_k + \Hmap^{-1} \bs g_k)}\right),
  \label{eq:qsnmap}
\end{equation}
which is a Gaussian with mean $\params_k - \Hmap^{-1} \bs g_k$ and
covariance matrix $\Hmap^{-1}$. Note that the scaling factor is not
necessary in~\eqref{eq:qsnmap}, since for proposals with MAP-based
Hessians, the scaling factors are constant and thus they cancel when
computing the acceptance probability $\alpha_k$ in
Algorithm~\ref{algorithm:M-H}. The SNMAP proposal is illustrated in
Figure~\ref{fig:proposals} (top right). Note that the SNMAP
proposal~\eqref{eq:qsnmap} can also be understood as a preconditioned
Langevin MCMC proposal~\cite{StuartVossWiberg04} with a MAP-based
Hessian preconditioner.

Avoiding SN's Hessian recomputation at each sample point results in
substantial computational savings, since, as will be made explicit in
Section \ref{subsec:lra}, computing the Hessian typically requires a
number of forward/adjoint PDE solves on the order of the effective
rank of (a properly preconditioned) Hessian of the data misfit term in
the negative log posterior. Once SNMAP has computed the Hessian at the
MAP point, the only cost per sample is a pair of forward/adjoint PDE
solves to compute the gradient. However, this may result in a
deterioration in the acceptance rate, since the Gaussian proposal
employs local gradient information but a global Hessian, and thus may
not fully capture local curvature information of the posterior if the
curvature is changing rapidly (as may happen in a highly nonlinear
parameter-to-observable map). However, the fact that the proposal
$q^{\SNMAP}$ changes less from sample to sample compared to $q^{\SN}$
can also have a positive effect on the acceptance probability and the
chain convergence.
In Section~\ref{sec:uq} we conduct numerical experiments on a specific
Bayesian inverse problem to assess whether this tradeoff is
profitable.

\subsection{Independence sampling with a MAP point-based Gaussian
  proposal (ISMAP)}
\label{subsec:MAPapprox}

As seen in the above SNMAP modification of the stochastic Newton MCMC
method, freezing the Hessian at the MAP point avoids Hessian
recomputation and results in substantial savings. However, the
gradient is still recomputed at each sample point, motivated by the
desire to construct a Gaussian proposal that captures some local
information, as well as the fact that the gradient is far cheaper to
compute than (a low rank-approximation-based) Hessian.

One can go one step further and shed the need to compute local
gradient information by defining an independence sampler that takes
the proposal to be a Gaussian centered at the MAP point, using the
Hessian computed at the MAP as the inverse covariance, and neglecting
the gradient (since it vanishes at the MAP). This method has been
suggested previously in the subsurface flow inversion literature
\cite{OliverReynoldsLiu08,OliverCunhaReynolds97}; here, we refer to it as {\em ISMAP}. Since
ISMAP, like SN and SNMAP, makes use of Hessian
information, a fair assessment of SNMAP vis-\'{a}-vis SN should include
comparisons to ISMAP as well. Therefore, we next provide a description of
the ISMAP proposal as well.

The proposal density $q^{\IS}$ is obtained by taking $\params_k$
in~\eqref{eq:local_q} as the MAP point $\map$ (which means that
$\bs{g}_k$ is zero) and, as with SNMAP, replacing the Hessian at
$\params_k$ with the Hessian evaluated at the MAP point, $\Hmap$. This
leads to
\begin{equation}
  q^{\IS}(\map,\proposal) \propto
  \exp
  \left(
  - \frac 12 \mip{\proposal - \map}{\Hmap (\proposal - \map)}\right),
  \label{eq:qis}
\end{equation}
which is a Gaussian with mean $\map$ and covariance
matrix $\Hmap^{-1}$. We note that the proposal $q^{\IS}$ is
independent of the current sample point, and thus does not change
during the sampling process. The ISMAP proposal is illustrated in
Figure~\ref{fig:proposals} (bottom).

We note that ISMAP not only avoids Hessian recomputation at each
sample point (as with SNMAP) but also avoids computing the gradient;
thus, its cost---once the MAP-based Hessian is determined---is
a forward PDE solve at each sample point.  However, this
additional approximation 
over SNMAP has the potential to lead to additional deterioration of
the acceptance rate. Note that one advantage of ISMAP is that, since
the proposal is constant, the samples can all be precomputed offline
or in parallel, after which they can be subjected (sequentially) to
the M-H accept/reject criterion in Algorithm~\ref{algorithm:M-H}.

Finally, we remark that if the posterior itself is a Gaussian,
the three Hessian-based methods described above collapse to the same
method. As such, they all sample from the true posterior with
probability 1 at every step, resulting in an acceptance rate of 100\%
and posterior samples that are independent
\cite{MartinWilcoxBursteddeEtAl12}.

\subsection{Relation to Newton's method for optimization}

Recall that the stochastic Newton MCMC method (in particular SN) uses,
as a proposal, the local quadratic approximation $\tilde J_k(\params)$
of the negative log posterior $J(\params)$ about the current sample
point $\params_k$. The minimizer of $\tilde J_k(\params)$ is given by
$
\params_k - \H_k^{-1}\bs g_k$, where $\H_k$ and $\bs g_k$ are the
Hessian and the gradient of $J(\params)$ evaluated at $\params_k$,
respectively. Note that $- \H_k^{-1}\bs g_k$ is the classical Newton
optimization step. A proposal point drawn from the local Gaussian
approximation of the posterior with mean $\params_k - \H_k^{-1}\bs
g_k$ and covariance $\H_k^{-1}$ is thus
\begin{align}
  \proposal = \params_k - \H_k^{-1}\bs g_k + \H_k^{-1/2} \gbf{\tilde
    n},\label{eq:V2sample}
\end{align}
where $\gbf{\tilde n} = \M^{-1/2} \gbf n$ is a random sample from a Gaussian 
with zero mean and identity covariance matrix in $\R^n_{\M}$, and
$\gbf n\in \R^n$ 
is a random sample from the standard normal density 
in $\R^n$. Iterating the stochastic Newton MCMC method  without
the random term amounts to the classical Newton method from nonlinear
optimization, which converges to the MAP point (or another stationary
point of $J(\cdot)$).

Since SNMAP reuses the Hessian at the MAP point (i.e., it is held
constant throughout the sampling process), proposal points are
computed as in \eqref{eq:V2sample}, but with $\H_k$ replaced by
the Hessian at the MAP point $\Hmap$, i.e.,
\begin{align}
  \proposal = \params_k - \Hmap^{-1}\bs g_k +
  \Hmap^{-1/2} \gbf{\tilde n},\label{eq:V3sample}
\end{align}
with $\gbf{\tilde n}$ as above.  We note that if the random term is
neglected, SNMAP reduces to an  $\Hmap$-preconditioned steepest 
descent method. 

For completeness, let us show how proposals from the independence
sampler with MAP-based Gaussian, ISMAP, are computed. With
$\gbf{\tilde n}$
defined as above, the proposed point is found as
\begin{align}
  \proposal = \map + \Hmap^{-1/2} \gbf{\tilde
    n}.
\label{eq:V1sample}
\end{align}
Note that the right hand side in \eqref{eq:V1sample} is
independent of $\params_k$, and hence the designation ``independence
sampler.''

\subsection{Efficient operations with the Hessian via low-rank
  approximation} 
\label{subsec:lra}

Up to this point, we have described the three Hessian-based MCMC
methods (SN, SNMAP, and ISMAP) 
in terms of the Hessian matrix of the negative log posterior. Indeed,
examination of the form of the three proposal densities
\eqref{eq:local_q}, \eqref{eq:qsnmap}, and \eqref{eq:qis}, as well as
the expressions for the samples from the proposals \eqref{eq:V2sample},
\eqref{eq:V3sample}, and \eqref{eq:V1sample}, reveals that the following
operations with
the Hessian are required: action of the Hessian on a vector; action of
the inverse Hessian on a vector; action of the inverse of the square
root of the Hessian on a vector; and determinant of the square root of
the Hessian (the determinant is required only for SN). 

Unfortunately, explicitly computing the Hessian requires as many
(linearized) forward PDE solves as there are parameters; for
large-scale problems, these computations are prohibitive. Thus, we
need efficient algorithms for the operations with the Hessian
summarized above. In this section, we briefly describe previous work
that employs low-rank approximations of the data misfit portion of the
Hessian, preconditioned by the
prior covariance, to execute all of the above operations with the
Hessian at a cost (measured in forward PDE solves) that is independent
of the parameter dimension \cite{FlathWilcoxAkcelikEtAl11,
  MartinWilcoxBursteddeEtAl12, Bui-ThanhGhattasMartinEtAl13}. The
discussion below is in terms of a generic Hessian, $\H$;
this can refer to the Hessian at any point in parameter space,
including the MAP point.

The Hessian of the negative log posterior $J(\params)$ in
\eqref{eq:cost} can be written as the sum of the Hessian of the data
misfit term, $\HM$, and the inverse of the prior
covariance~$\prcov^{-1}$.  If we consider a decomposition of the prior
such that $\prcov = \L\adjMacroMM{\L}$, then
\begin{equation}\label{eq:H}
  \H = \HM + \prcov^{-1} =  \HM + {\adjMacroMMINV{\L}} \L^{-1}  =
  \adjMacroMMINV{\L} ( \adjMacroMM{\L} \HM \L + \I) \L^{-1}.
\end{equation}
Here, the  data misfit Hessian $\HM$ is given by
\begin{align*}
\HM :=  \Fadj\matrix{\Gamma}_{\text{noise}}^{-1}\F 
\; + \; \text{second order terms} 
\end{align*}
where $\bs{F}$
is the Jacobian matrix of the parameter-to-observable map
$\bs{f}(\bs{m})$, $\Fadj := \M^{-1}\F^T$ is its (properly weighted)
adjoint, and the second order terms involve second derivatives of
$\bs{f}(\bs{m})$ with respect to $\bs{m}$. Notwithstanding the form of
the second order terms, the expression above suggests that the Hessian
of the data misfit involves the solution of linearized forward and
adjoint PDE problems. This will be seen explicitly for the target ice
sheet inverse problem described in Section \ref{sec:app}.

We begin by describing the computation of the application of the inverse
Hessian to a vector in order to compute the Newton step $\H^{-1} \bs
g$. From~\eqref{eq:H}, we obtain
\begin{equation}\label{eq:Hinvg}
  \H^{-1} \bs g = \L ( \adjMacroMM{\L} \HM \L + \I)^{-1}
  \adjMacroMM{\L} \bs g,
\end{equation}
and thus we require the inverse of $( \adjMacroMM{\L} \HM \L + \I)$.
Since for ill-posed inverse problems, observations typically inform
only a limited number of eigenvectors of the parameter field, the spectrum of
the data misfit Hessian often decays rapidly (see for example,
\cite{Bui-ThanhGhattas12, Bui-ThanhGhattas12a, Bui-ThanhGhattas12f}
for the inverse scattering case). In addition, the prior is often
smoothing, in which case left and right preconditioning of the data
misfit Hessian by the square root of the prior, $\L$, enhances the
decay of the eigenvalues. Thus, the prior-preconditioned data misfit Hessian, $\adjMacroMM{\L} \HM \L$,
can typically be well approximated by a low rank matrix, and this can
be exploited to enable efficient computations with the Hessian.  To
construct the low-rank approximation of the prior-preconditioned data
misfit Hessian, we seek a matrix-free method (since $\H$ cannot be
formed explicitly) that requires just Hessian-vector products;
crucially, the number of Hessian-vector products must be of the order
of the effective rank, $r$, of the prior-preconditioned data misfit
Hessian, as opposed to the parameter dimension, $n$. Note that each
Hessian-vector product can be formed efficiently at the cost of a
single pair of linearized forward/adjoint PDE solves (this will be
seen explicitly for the ice sheet flow problem in Section
\ref{subsec:app:gH}).

The Lanczos eigenvalue algorithm meets the requirements outlined
above, and we use it to construct an $r$-dimensional low-rank
approximation for the prior-preconditioned data misfit Hessian, i.e.,
$\adjMacroMM{\L} \HM \L \approx \Vr \matrix{\Lambda}_r \Vadj_r$, where
$\Vr \in \mathbb{R}^{n\times r}$ contains $r$ eigenvectors of the
prior-preconditioned data misfit Hessian corresponding to the $r$
largest eigenvalues $\lambda_i, i=1, \ldots,r$, $\matrix{\Lambda}_r =
\diag (\lambda_1, \hdots, \lambda_r) \in \mathbb{R}^{r \times r}$, and
$\Vadj_r \in \R^{r\times n}$ denotes the adjoint defined
in~\eqref{eq:EM_adj}.  The rank $r$ approximation can typically be
formed in a number of Hessian-vector products that is slightly larger
than $r$, which amounts to approximately $r$ forward/adjoint pairs of
linearized PDE solves all containing the same PDE operator or its
adjoint (this presents an opportunity to employ an effective PDE
preconditioner, since it will be amortized over $r$ PDE
forward/adjoint solves.)
Once the low-rank approximation has been constructed, the product
of this approximate Hessian with a vector can then be formed by
successively applying $\Vr$ and $\Vadj_r$ to vectors, each application 
amounting to $r$ inner products. The cost of this linear algebra is
negligible relative to the PDE solves needed to form the low-rank
approximation. 

Moreover, using the Sherman-Morrison-Woodbury
formula~\cite{GolubVan96} in combination with expressing the
prior-preconditioned data misfit Hessian as the sum of a low rank term
and a reminder, we can write the inverse Hessian as
\begin{equation}
\label{eq:eigen_error}
( \adjMacroMM{\L} \HM \L + \I)^{-1}=
\matrix{I}-\Vr \Dr \Vadj_r +
\mathcal{O}\left(\sum_{i=r+1}^{n} \frac{\lambda_i}{\lambda_i + 1}\right),
 \end{equation}
where $\matrix{D}_r :=\diag(\lambda_1/(\lambda_1+1), \hdots,
\lambda_r/(\lambda_r+1)) \in \mathbb{R}^{r\times r}$. As can be seen
from the form of the remainder term above, to obtain an accurate low
rank approximation of $\H^{-1}$, we can neglect eigenvectors
corresponding to eigenvalues that are small compared to~$1$. Therefore,
\begin{equation}\label{eq:Hrinvg}
  \H^{-1} \bs g \approx \L ( \matrix{I}-\Vr \Dr \Vadj_r)
  \adjMacroMM{\L} \bs g =  \L \big\{ \Vr \big[ (\matrix{\Lambda}_r +
    \Ir)^{-1} - \Ir \big]  \Vr^{\diamond} + \bs I \big\}
  \adjMacroMM{\L} \bs g.
\end{equation}
The expression on the right side of~\eqref{eq:Hrinvg} can be used to
efficiently apply the square-root inverse Hessian to a vector $\bs x$,
as needed for drawing samples from a Gaussian distribution with
covariance $\H^{-1}$. Namely,
\begin{equation}\label{eq:Hrinvsqrtx}
  \H^{-1/2} \bs x \approx \L \big\{ \Vr \big[ (\matrix{\Lambda}_r +
    \Ir)^{-1/2} - \Ir \big] \Vadj_r + \bs I \big\}\bs x.
\end{equation}
By a direct computation using the adjoint definitions \eqref{eq:MM_adj}
and \eqref{eq:EM_adj}, it can be verified that $\H^{-1}\bs x =
\H^{-1/2}\adjMacroMM{(\H^{-1/2})}\bs x$.
Finally, the determinant of the square-root Hessian can be
computed efficiently from
\begin{align}
\det (\Hpost^{1/2}) &= (\det \Lpr)^{-1} \prod_{i=1}^r (\lambda_i +
1)^{1/2} \label{low_rank_determinant}.
\end{align}

In summary, once the low-rank approximation of the data misfit Hessian
has been constructed, all of the operations with the Hessian described
above (and required by the three Hessian-based methods) can be carried
out using only inner products and vector sums, without recourse to PDE
solves. These linear algebra operations are negligible relative to the
PDE solves needed for the low-rank approximation, and thus the
dominant cost of these methods is $O(r)$ forward/adjoint PDE solves
needed for the low-rank approximation. As mentioned above, for
ill-posed inverse problems (including the ice sheet flow inverse
problem studied below), the prior-preconditioned data misfit Hessian
is a compact operator with rapidly-decaying eigenvalues, so that $r
\ll n$. Moreover, when the dominant eigenvectors of the
prior-preconditioned data misfit Hessian are spatially smooth,
$r$ is independent of the parameter dimension $n$ and the observation
dimension $q$.

\subsection{Comparison of computational cost of ISMAP, SNMAP, and SN}

The stochastic Newton MCMC methods and the independence sampler with
MAP-based Gaussian all use the low-rank approximation and fast
operations with the Hessian described in the previous
section. However, they differ markedly in how frequently they
recompute the low-rank approximation of the prior-preconditioned data
misfit Hessian, which as mentioned above is by far the dominant cost
relative to the linear algebra.

Let us now characterize the cost per MCMC sample for each of the three
Hessian-based methods described in Section~\ref{sec:proposals},
measured in number of (forward or adjoint) PDE solves.  The
independence sampling method (ISMAP) requires just a single evaluation of
the parameter-to-observable map per sample, which amounts to a single
(nonlinear) forward PDE solve per sample.  The stochastic Newton MCMC
method with dynamically changing Hessian (SN) requires for each sample
a nonlinear forward PDE solve, a (linear) adjoint PDE solve for the
gradient computation, and approximately $2r$ linearized PDE solves to
construct the rank $r$ approximation of the prior-preconditioned data
misfit Hessian.
Finally, the cost per sample for stochastic Newton MCMC with MAP-based
Hessian (SNMAP) is one nonlinear forward PDE solve and 
one adjoint PDE solve, since SNMAP recomputes the gradient at each sample
point. Depending on whether the forward problem 
is linear or nonlinear and stationary or time dependent, and depending
on whether the linearized PDEs are solved by direct factorization
(which permits reuse of the factors within the low-rank approximation)
or iteratively (which permits reuse of only the preconditioner), the
number of PDE solves per sample translates differently into
computational time per sample. Thus, the metric we use to compare the
performance of these three Hessian-based methods to each other in
Section \ref{sec:uq} is the number of linearized PDE solves required
by each method.

\section{Application to the inversion of basal boundary conditions in
  ice flow problems}
\label{sec:app}

In the remainder of this paper, we apply the methods discussed in
Section~\ref{sec:proposals} to an inverse problem in ice dynamics, in
which we seek to find a statistical description of the uncertain basal
sliding coefficient field from pointwise velocity observations at the
surface of the moving mass of ice. In this section, we summarize the
physics describing the dynamics of ice flows, present the
two-dimensional problem used to exercise our methods, and the prior
distribution and the likelihood for the Bayesian inverse problem.
We also give expressions of the gradient and the Hessian-vector
product of the negative log posterior function using adjoint ice
flow equations and describe the discretization of these equations.

\subsection{The dynamics of ice flow}
\label{subsec:app:fwdmodel}

\begin{figure}
 \begin{center}
   \begin{tikzpicture}
     \node (img1)
           {\includegraphics[width=0.85\columnwidth]{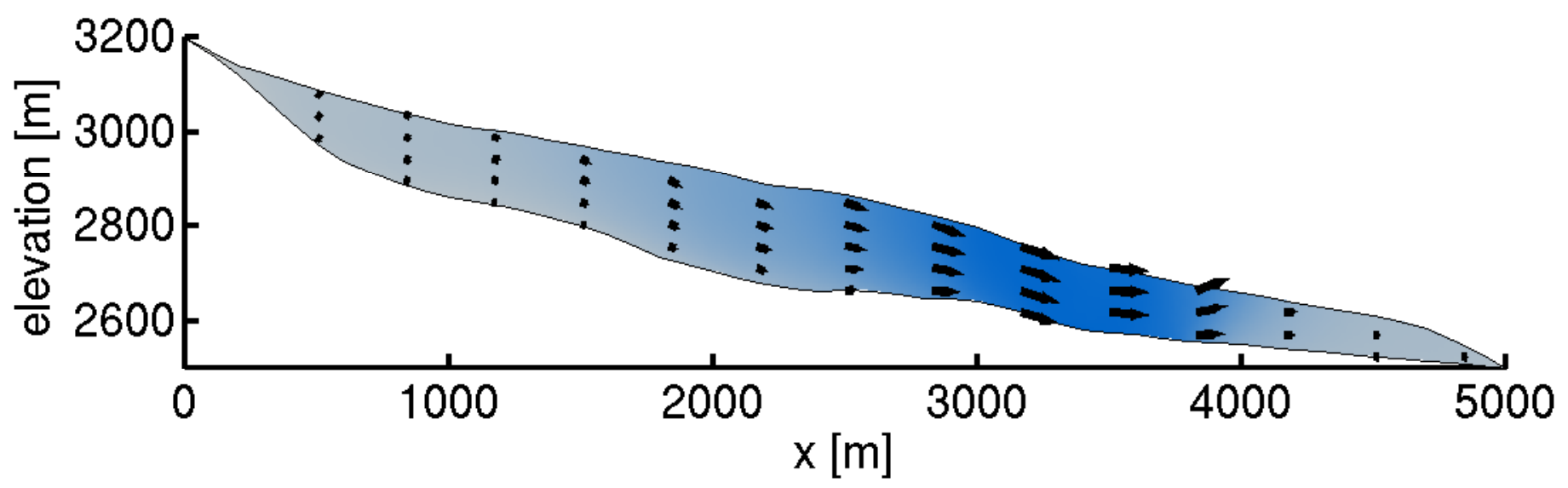}};
           \node (img1) at (0.15\columnwidth, 0.12\columnwidth)
                 {\includegraphics[width=0.5\columnwidth]{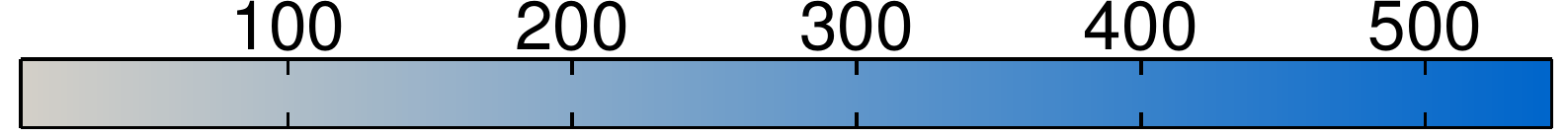}};
                 \node at (2,0.15) {$\Gamma_{\!\text{\scriptsize$t$}}$};
                 \node at (-1.5,-0.15) {$\Gamma_{\!\text{\scriptsize$b$}}$};
   \end{tikzpicture}
 \end{center}
  \caption{The longitudinal profile of Haut Glacier d'Arolla from the
    ISMIP-HOM benchmark collection
    \cite{PattynPerichonAschwandenEtAl08}. This profile follows
    a flowline of 5 km length with a grid spacing of 100 m. The arrows
    represent the flow field obtained by solving~\eqref{eq:stokes}
    with the basal sliding coefficient field given by~\eqref{eq:beta}.}
  \label{fig:geometry}
\end{figure}

We model the flow of ice as a non-Newtonian, viscous, incompressible,
isothermal fluid~\cite{Hutter83, Marshall05, Paterson94, GreveBlatter09}. The balance
of mass and linear momentum in a domain $\Omega\subset \mathbb{R}^d$
of dimension $d=2$ or $d=3$ state that
\begin{subequations}\label{eq:stokes}
\begin{alignat}{2}
  \gbf{\nabla} \cdot \gbf{u} &= 0 & &\text{ in } \Omega, \label{eq:model:mass}\\
  - \gbf{\nabla} \cdot \gbf \sigma_{\!\gbf u} &= \rho \gbf{g} \ & &\text{ in }
  \Omega, \label{eq:model:momentum}
\end{alignat}
where $\gbf u$ denotes the velocity vector, $\gbf
\sigma_{\!\gbf u}$ the stress tensor, $\rho$ the density of the ice,
and $\gbf g$ gravity. The stress, $\gbf \sigma_{\!\gbf u}$, can be
decomposed as $\gbf \sigma_{\!\gbf u} = \gbf \tau_{\!\gbf u} - \gbf{I}
p$, where $\gbf \tau_{\!\gbf u} $ is the deviatoric stress tensor, $p$
the pressure, and $\gbf{I}$ the unit tensor. We employ a constitutive
law for ice that relates stress and strain rate tensors by Glen's flow
law~\cite{Glen55},
\begin{equation}
  \gbf \tau_{\!\gbf u} = 2 \eta(\gbf u) \edot_{\gbf{u}}, \;
  \text{with } \eta(\gbf u)
  =  \frac12 A^{-\frac1n} \;
  \secinve^{\frac{1-n}{2n}},\label{eq:glenslaw}
\end{equation}
where $\eta$ is the effective viscosity, $\edot_{\gbf{u}} = \frac12
(\gbf{\nabla u} + \gbf{\nabla u}^T)$ the strain rate tensor, $\secinve
= \frac12 \mathrm{tr}(\edot_{\gbf{u}}^2)$ its second invariant, $n\ge
1$ Glen's flow law exponent, and $A$ the temperature-dependent flow
rate factor (here taken as constant in isothermal ice).

At the base $\Gamma_{\!\mbox{\scriptsize \mbox{b}}}$ of the ice sheet,
one commonly assumes non-penetrating normal boundary conditions and a
linear sliding law for the tangential components
i.e.~\cite{Paterson94}
\begin{equation}
  \gbf{u}\cdot \gbf{n} = 0, \,\,\, \gbf T \gbf{\sigma}_{\!\gbf{u}}
  \gbf{n} + \exp(\beta) \gbf T\gbf{u} = \gbf{0},\label{eq:sliding}
\end{equation}
where $\beta=\beta(\bs x)$ is the log basal sliding coefficient field,
and $\gbf T := \gbf I - \gbf n \otimes \gbf n$ the projection onto
the tangential plane. Here, ``$\otimes$'' represents the tensor (or
outer) product defined by $(\gbf a \otimes \gbf b)\gbf c = \gbf a \gbf
b \cdot \gbf c$,  $\gbf n$ is the outward normal vector, and $\gbf I$ is
the second order unit tensor. Together with appropriate boundary conditions on
$\partial\Omega\setminus \Gamma_b$,~\eqref{eq:stokes} represents an
accepted model for the flow of ice sheets and glaciers.
Note that the Robin coefficient field
$\exp(\beta)$, which relates tangential velocity to tangential
traction, subsumes several complex physical phenomena such as the
frictional behavior of the ice sheet, the roughness of the bedrock and
hydrological phenomena. It does not itself represent a physical
parameter and is highly uncertain. Our target is to infer the log
sliding coefficient field $\beta$, which in the following we simply refer to
as sliding coefficient field, within a Bayesian inversion approach.
In the next section, we specify the ice flow model problem
used to study the efficiency of our algorithms and to interpret
results of Bayesian inversions.

\subsection{The Arolla test problem}
We use a two-dimensional test problem taken from the Ice Sheet Model
Intercomparison Project for Higher-Order Ice Sheet Models (ISMIP-HOM)
benchmark study~\cite{PattynPerichonAschwandenEtAl08}. The domain
$\Omega$, which is based on data from the Haut Glacier d'Arolla is
shown in Figure~\ref{fig:geometry}. Together with the basal boundary
condition~\eqref{eq:sliding}, on the top boundary
$\Gamma_{\!\mbox{\scriptsize $t$}}$ we assume the traction-free condition
\begin{alignat}{2}
  \gbf{\sigma}_{\!\gbf{u}} \gbf{n} &= \gbf{0}  & &\text{ on }
  \Gamma_{\!\mbox{\scriptsize $t$}}.  \label{eq:model:bctraction}
\end{alignat}
\end{subequations}
The driving force in the Stokes equations~\eqref{eq:stokes} is the
gravity $\rho{\gbf g} = (0, -\rho g \cos \theta)$, where $\rho = 910$
kg/m$^3$ is the ice density, and $g =9.81$ m/s$^2$ is the
gravitational constant.  The Glen's flow-law exponent parameter is $n
= 3$, and the rate factor is assumed constant as $A = 10^{-16}$
Pa$^{-n}$a$^{-1}$, where ``Pa'' and ``a'' are units of Pascals and
years, respectively~\cite{PattynPerichonAschwandenEtAl08}.

As reference basal sliding coefficient field, which is also used to generate
synthetic observations as described in the next section, we choose
\begin{equation}\label{eq:beta}
  \beta_{\scriptsize \text{true}}(x) = \ln
  \left\{
  \begin{array}{ll}
    1000+1000\sin\left(\frac{2\pi x}{5000}\right)  & \mbox{if } 0 \leq x < 3750, \\
    1000\left(16 - \frac{x}{250}\right)         & \mbox{if } 3750 \leq x < 4000, \\
    1000                                        & \mbox{if } 4000 \leq x < 5000.
  \end{array}
  \right.
\end{equation}
The flow field corresponding to the basal sliding coefficient
field~\eqref{eq:beta} are shown in Figure~\ref{fig:geometry}.

\subsection{The likelihood}
\label{subsec:app:like}
The likelihood function expresses the probability that a candidate set
of parameters reproduces the observations $\yobs$.  To specify the
likelihood function, we denote by $\uu(\beta)$ the solution of the
Stokes equation with basal sliding coefficient field $\beta(\bs x)$,
and by $\mathcal B$ the observation operator, which restricts the flow
solution to ten measurement points on the right of the top surface
$\Gamma_{\!\mbox{\scriptsize $t$}}$, i.e., lower part of the glacier,
with $x$-coordinates uniformly distributed in $[2500,5000]$.
Thus, the parameter-to-observable map is $\ff(\beta)=
\mathcal B\uu(\beta)$.

The observations $\yobs$ are synthetically generated by solving the ice
flow Stokes equations \eqref{eq:stokes} with basal sliding coefficient
field~$\beta_{\scriptsize \text{true}}$ as specified in \eqref{eq:beta},
restricting the resulting flow solution
$\uu(\beta_{\scriptsize\text{true}})$ using the observation operator
and adding additive Gaussian noise $\bs \epsilon$, i.e.,
$\yobs=\mathcal B\uu(\beta_{\scriptsize\text{true}}) + \bs \epsilon$.
Each component of the noise vector $\bs \epsilon$ is i.i.d.\ with
standard deviation $\bar\sigma_{\scriptsize{\text{noise}}}$ for the
horizontal flow components and with
$\bar{\bar\sigma}_{\scriptsize{\text{noise}}}$ for the vertical flow
components. Adding noise mitigates the ``inverse crime,'' which
occurs when synthetic observations are used in an inversion and the
same numerical method is employed in both, the synthetization of the
observations and in the inverse problem
solution~\cite{KaipioSomersalo05}.  The likelihood function is then
given by
\begin{equation}
\like(\yobs | \dbeta )
\propto
 \exp \biggl[-\half
\nor{{\mathcal B}\uu(\beta) - \yobs}_{\ncov^{-1}}^2\biggr], 
\end{equation}
where the noise covariance matrix $\ncov$ is diagonal with the entries
${\bar\sigma}_{\scriptsize{\text{noise}}}$ and
$\bar{\bar\sigma}_{\scriptsize{\text{noise}}}$ for the horizontal and
vertical components, respectively. To understand the effect of the
noise level on the performance of the three Hessian-based sampling
methods and on the uncertainty in the reconstruction,
we consider two problems based on the noise level in the
observations:
\begin{itemize}

\item {\bf Problem 1:} $\bar\sigma_{\scriptsize{\text{noise}}} = 62$
  for the horizontal flow components and with
  $\bar{\bar\sigma}_{\scriptsize{\text{noise}}} = 10$ for the vertical
  flow components;

\item {\bf Problem 2:} $\bar\sigma_{\scriptsize{\text{noise}}} = 18$
  for the horizontal flow components and with
  $\bar{\bar\sigma}_{\scriptsize{\text{noise}}} = 3$ for the vertical
  flow components.
\end{itemize}

\subsection{The choice of prior}
\label{subsec:app:prior}
We specify the Gaussian prior by giving its mean~$\beta_0$ and its
covariance via the elliptic operator $\mc{A}$ discussed in
Section~\ref{sec:prelim}. Since the bottom surface of the Arolla
geometry is a ``curved'' surface, the prior is defined in terms of the
surface Laplacian (also called the Laplace-Beltrami operator).  Using
the projection $\gbf T$ onto the tangential plane as defined above, $
\gbf{\nabla}_{\Gamma_{\!b}} = \gbf T \gbf \nabla$ is the tangential
gradient, $\gbf \nabla_{\Gamma_{\!b}} \cdot$ is the tangential
divergence, and $\gbf{\nabla}_{\Gamma_{\!b}} \cdot \gbf
\nabla_{\Gamma_{\!b}}$ is the Laplace-Beltrami
operator~\cite{BonitoNochettoPauletti10, Demlow09, Dziuk88}.  Thus, we
define $\mc{A}$ as the differential operator
\begin{subequations}\label{eq:priorop}
\begin{align}
- \gbf \nabla_{\Gamma_{\!b}} \cdot (a \gbf \nabla_{\Gamma_{\!b}}
\beta) + b \beta &= s \quad \text{in } \Gamma_{\!b},\label{eq:priorop1}\\
 (a \gbf \nabla_{\Gamma_{\!b}} \beta) \cdot \gbf \nu &= 0 \quad \text{on }
\partial\Gamma_{\!b},\label{eq:priorop2}
\end{align}
\end{subequations}
where $\gbf \nu$ denotes the outward unit normal on $\partial
\Gamma_{\!b}$. The finite-dimensional representation of the prior
inverse is $\prcov^{-1} = \M^{-1}\K$, where $\M$ and $\K$ are the
corresponding surface mass and surface stiffness matrices,
respectively.  In our model problems, we use the parameters $a = 10^{-2}$ and $b
= 10^2$. With these parameters, the standard deviation of the Green's
function corresponding to the prior (i.e., the correlation length) is
roughly $5\%$ of the total length of the glacier.

\subsection{Gradient and Hessian of the negative log posterior}
\label{subsec:app:gH}

The Hessian-based sampling methods presented in
Section~\ref{sec:proposals} rely on the availability of gradients and
Hessian-vector products of the negative log posterior. The derivation
of these derivatives is complicated by the fact that the
parameter-to-observable map involves the solution of the ice flow
equations.  In this section, we give expressions for the efficient
computation of gradients and Hessian-vector products using adjoint
equations. For a more detailed presentation of derivative computation
using adjoints we refer to the PDE-constrained optimization
monographs~\cite{BorziSchulz12,HinzePinnauUlbrichEtAl09,Troltzsch10}
and to~\cite{PetraZhuStadlerEtAl12} for the ice flow dynamics setting.

The gradient of the negative log posterior can be found by requiring
that variations of a Lagrangian function with respect to the forward
velocity and pressure $(\gbf u,p)$ and an adjoint velocity and
pressure $(\gbf v,q)$ vanish. Variations with respect to $\beta$ then
result in the following strong form of the gradient $\mathcal{G}$:
\begin{equation}\label{eq:gradient}
  \mc{G}(\beta):= \exp(\beta) \gbf T \uu \cdot \gbf T \vv + \mc{A}(\beta-\beta_0).
\end{equation}
Here, the velocity $\uu$ is obtained by solving the {\it forward
  Stokes problem}~\eqref{eq:stokes} for given $\beta$, and the adjoint
velocity $\vv$ is obtained by solving the following {\it adjoint
  Stokes problem} for given $\beta$ and for $\uu$
satisfying~\eqref{eq:stokes}:
\begin{subequations}\label{eq:adjoint}
\begin{alignat}{2}
  \gbf{\nabla} \cdot \gbf{v} &= 0  &\; &  \quad \text{ in }
  \Omega, \label{eq:KKT:adj:1} \\
  - \gbf{\nabla} \cdot \gbf{\sigma}_{\!\gbf{v}} &=
  \gbf{0} & & \quad \text{ in }
  \Omega,\label{eq:KKT:adj:2}\\ \gbf{\sigma}_{\!\gbf{v}} \gbf{n} &=
  - \mathcal{B}^*\Gamma^{-1}_{\text{noise}} (\mathcal{B}\uu - \yobs)&\; &
  \quad \text{ on } \Gamma_{\!\mbox{\tiny}
    t}, \label{eq:KKT:adj:3}\\ \gbf{v}\cdot \gbf{n} = 0, \; \gbf
  T\gbf{\sigma}_{\!\gbf{v}} \gbf{n} + \exp(\beta) \gbf T\gbf{v} & =
  \gbf{0} &\; & \quad \text{ on } \Gamma_{\!\mbox{\tiny}
    b}, \label{eq:KKT:adj:4}
\end{alignat}
\end{subequations}
where the adjoint stress $\gbf{\sigma}_{\!\gbf{v}}$ is given by
\begin{displaymath}
\gbf{\sigma}_{\!\gbf{v}} := 2\eta(\gbf{u})
\, \bigl(\mathsf{I} + \frac{1-{n}}{{n}} \frac{\edot_{\gbf{u}}\otimes
  \edot_{\gbf{u}}}{\edot_{\gbf{u}} :\,
  \edot_{\gbf{u}}}\bigr)\edot_{\gbf{v}} -\gbf{I} q,
\end{displaymath}
and  $\mathsf{I}$ is the fourth order identity tensor.

The action of the Hessian operator evaluated at a sliding coefficient
field~$\beta$ onto a direction~$\hat \beta$ is given by
\begin{align}
  \mathcal{H}&(\beta)(\hat \beta) := \mc{A}\hat \beta + \exp(\beta) (
  \hat \beta \gbf T\uu \cdot \gbf T\gbf{v} +
             \gbf T\gbf{\hat u} \cdot \gbf T{\vv} +
             \gbf T\uu \cdot \gbf T\gbf{\hat v}), \label{eq:applyH}
\end{align}
where the
{\it incremental forward velocity/pressure} $(\gbf{\hat u}, \hat
p)$ satisfy  the {\it incremental forward Stokes problem},

\begin{subequations}\label{eq:incremental-forward}
\begin{alignat}{2}
  \gbf{\nabla} \cdot \gbf{\hat u} & = \,0 & &  \quad \text{ in } \Omega,\\
  - \gbf{\nabla} \cdot \gbf{\sigma}_{\!\gbf{\hat u}} &= \gbf{0} &
           &  \quad \text{ in } \Omega,\label{eq:incfwd2}\\
  \gbf{\sigma}_{\!\gbf{\hat u}} \gbf{n} & = \gbf{0} &
  &  \quad \text{ on }\Gamma_{\!\mbox{\tiny} t},\label{eq:incfwd3}\\
   \gbf{\hat u}\cdot \gbf{n} =  0, \; \gbf
   T\gbf{\sigma}_{\!\gbf{\hat u}} \gbf{n} +\exp(\beta) \gbf T
  \gbf{\hat u} & = - \hat \beta  \exp(\beta) \gbf T \uu  & &
  \quad \text{ on } \Gamma_{\!\mbox{\tiny}
    b}, \label{eq:incfwd4}
\end{alignat}
\end{subequations}
with $\gbf{\sigma}_{\!\gbf{\hat u}} := 2 \eta(\gbf{u})\,
\bigl(\mathsf{I} + \frac{1-n}{n} \frac{\edot_{\gbf{u}}\otimes
  \edot_{\gbf{u}}}{\edot_{\gbf{u}} : \,
  \edot_{\gbf{u}}}\bigr)\edot_{\gbf{\hat u}}-\gbf{I} \hat p$, and the
     {\em incremental adjoint velocity/pressure} $(\hat{\gbf v}, \hat
     q)$ satisfy the {\it incremental adjoint Stokes problem},
\begin{subequations}\label{eq:incremental-adjoint}\\
\begin{alignat}{2}
  \gbf{\nabla} \cdot \gbf{\hat v} & = \,0  &\:
  &\text{ in } \Omega, \label{eq:incadj1}\\
  - \gbf{\nabla} \cdot \gbf{\sigma}_{\!\gbf{\hat v}} &=
  - \gbf{\nabla} \cdot \tau_{\!\gbf{\hat u}} & &\text{ in } \Omega,\label{eq:incadj2}\\
  \gbf{\sigma}_{\!\gbf{\hat v}} \gbf{n} & =
  -\mathcal{B}^*\Gamma^{-1}_{\text{noise}}\mathcal{B}\gbf{\hat u} - \tau_{\!\gbf{\hat u}} &\: &\text{ on }
  \Gamma_{\!\mbox{\tiny} t},\label{eq:incadj3}\\
  \gbf{\hat v}\cdot \gbf{n} = 0, \; \gbf T\gbf{\sigma}_{\!\gbf{\hat v}}
  \gbf{n} + \exp(\beta) \gbf T \gbf{\hat v} &= - \gbf T \tau_{\!\gbf{\hat u}}  \gbf{n}& & \text{ on }
  \Gamma_{\!\mbox{\tiny} b},\label{eq:incadj4}
\end{alignat}
\end{subequations}
with $\gbf{\sigma}_{\!\gbf{\hat v}} := 2 \eta(\gbf{u}) \,
\bigl(\mathsf{I} + \frac{1-n}{n} \frac{\edot_{\gbf{u}}\otimes
  \edot_{\gbf{u}}}{\edot_{\gbf{u}} : \,
  \edot_{\gbf{u}}}\bigr)\edot_{\gbf{\hat v}} -\gbf{I}\hat q$, and
$\tau_{\!\gbf{\hat u}} = 2 \eta(\uu) \mathsf\Psi \edot_{\gbf{\hat u}}$,
where
\begin{displaymath}
  \mathsf \Psi = (1 + \frac{1-n}{n} \edot_{\gbf{u}}
  : \,\edot_{\gbf{u}})\mathsf{I} +
  \frac{1-n}{n} \biggr[ \frac{\edot_{\gbf{u}}\otimes
    \edot_{\gbf{u}}}{\edot_{\gbf{u}} : \,\edot_{\gbf{u}}} +
  2 \frac{\edot_{\gbf{u}}\otimes
    \edot_{\gbf{v}}}{\edot_{\gbf{u}} : \,\edot_{\gbf{u}}} +
   \frac{1-3n}{n} \frac{\edot_{\gbf{u}}\otimes
    \edot_{\gbf{u}}}{(\edot_{\gbf{u}} : \,\edot_{\gbf{u}})^2} \biggl].
\end{displaymath}
In these expressions, $\edot_{\gbf{\hat u}}$ and $\edot_{\gbf{\hat
    v}}$ are defined analogously to $\edot_{\gbf{u}}$
and~$\edot_{\gbf{v}}$.

To summarize, the computational cost (measured in the number of
linearized Stokes solves, which represent the dominant cost) of the
gradient evaluation is $n_{\text{ls}}$ forward linearized Stokes
solves for the nonlinear forward problem \eqref{eq:stokes} (where
$n_{\text{ls}}$ is the number of Newton iterations required by the
nonlinear solver to converge), and one linear adjoint solve for
\eqref{eq:adjoint}. Each computation of the Hessian-vector product
\eqref{eq:applyH} requires two linearized Stokes solves, namely the
solution of \eqref{eq:incremental-forward} and
\eqref{eq:incremental-adjoint}.

\subsection{Discretization and solvers}\label{subsec:app_solvers}
We discretize the domain $\Omega$ with 260 triangular mesh elements,
and use Taylor-Hood finite elements (i.e., linear elements for
pressure and quadratic elements for the velocity components which
leads to 4714 degrees of freedom for the velocity field and 659 for
the pressure) for the forward and adjoint Stokes problems as well as
their incremental counterparts. The uncertain sliding coefficient
field $\beta$ is discretized using linear elements with 139 unknowns,
i.e., parameters for the inverse problem. We ensure that the state and
parameter fields are sufficiently resolved by comparing the solutions
computed on different meshes.  All Stokes systems are solved using a
direct factorization method. The cost of the Stokes matrix
factorization is amortized across the adjoint solve and the
incremental forward and adjoint solves in all CG iterations needed in
each Newton iteration;
when the factorization is available, only
triangular solutions are required at each CG iteration, gradient
computation or the application of the Hessian to a vector.

\section{Performance of algorithms}
\label{sec:uq}
The primary goal of this section is to compare the performance of the
sampling methods presented in Section~\ref{sec:proposals} for the
Bayesian inverse problem described in Section~\ref{sec:app}.  We start
with a discussion on the computation of the MAP point in
Section~\ref{subsec:MAP}, and study the approximation of
prior-preconditioned data misfit Hessians---and thus covariance
matrices---using low rank ideas (Section~\ref{subsec:lowrank}). In
Section~\ref{subsec:MCMCcomparison}, we present a systematic
comparison of the three Hessian-based sampling methods (ISMAP, SN, and
SNMAP) presented in Section~\ref{sec:proposals}.

\subsection{Computation of the MAP point}
\label{subsec:MAP}
For the computation of the MAP point, we apply an adjoint-based
inexact Newton method to solve the nonlinear least-squares
optimization problem~\eqref{eq:posterior_V}. Starting with an initial
guess for the basal sliding coefficient field $\beta$ (we use the
prior mean $\beta_0 \equiv \ln (1000)$), Newton's method iteratively
updates this parameter based on successive quadratic approximations of
the negative log posterior functional $J(\cdot)$, using the
expressions for the first and second derivatives presented in
Section~\ref{subsec:app:gH}. Since the conjugate gradient method is
used to solve the Newton linearization, the method does not require
assembled Hessian matrices but only Hessian-vector products. For a
more complete presentation of this optimization method to compute MAP
points for ice sheet model problems, we refer
to~\cite{PetraZhuStadlerEtAl12}.

\begin{center}
\begin{figure}[t]
  \includegraphics[width=0.49\columnwidth]{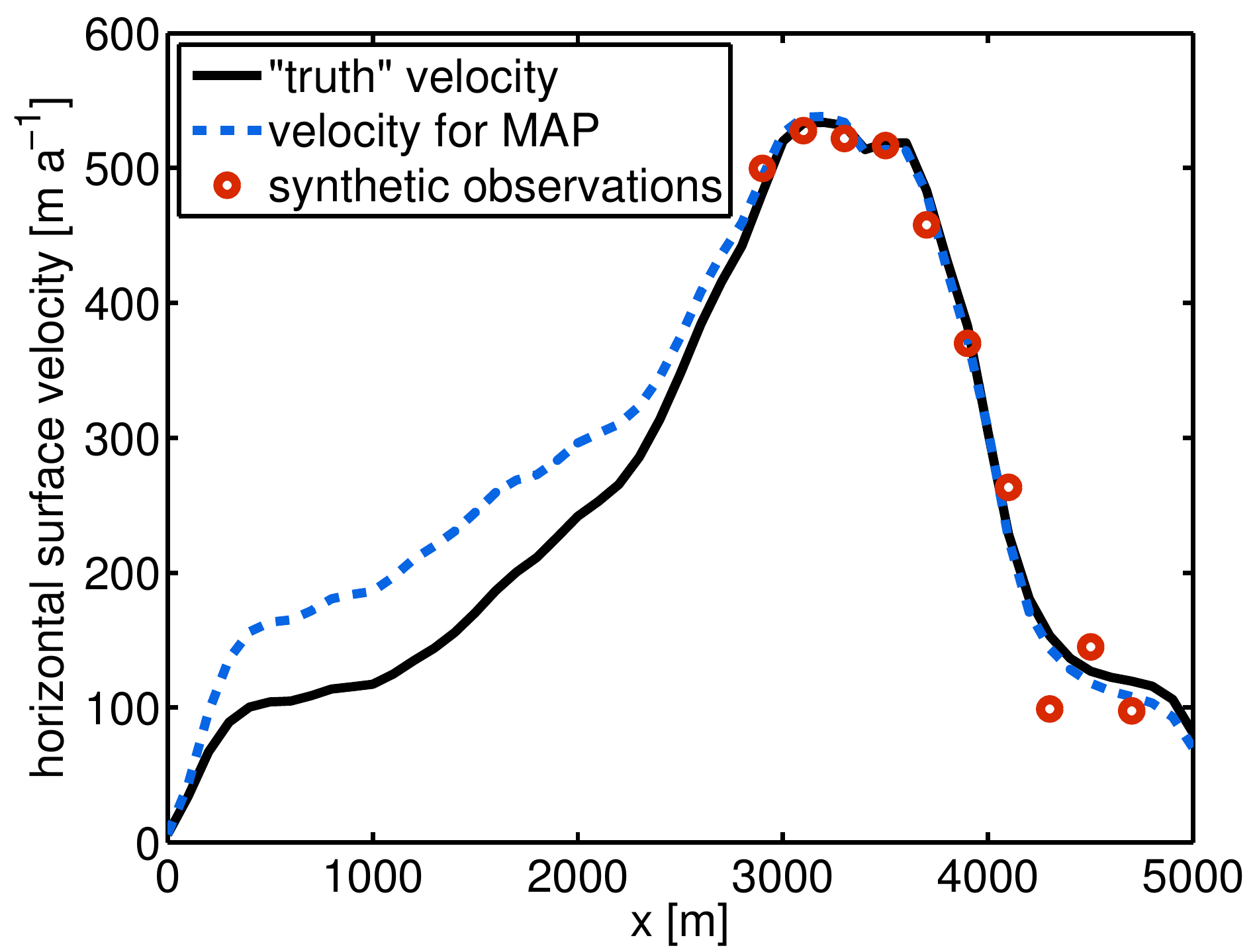}
  \includegraphics[width=0.465\columnwidth]{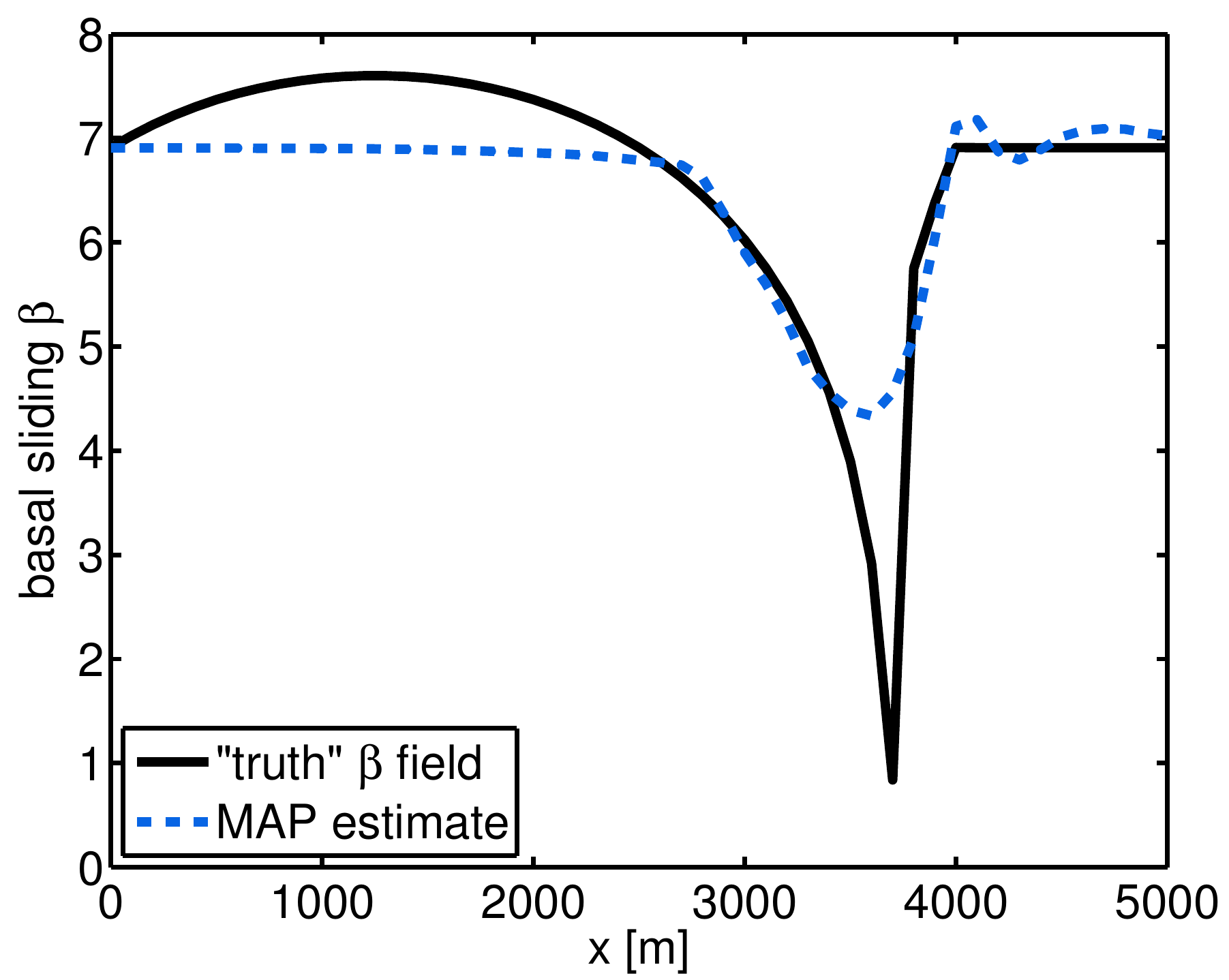}
  \caption{Left: The horizontal surface velocity obtained by
    solving the forward problem using
    the ``truth'' sliding coefficient field (solid line) and the synthetic pointwise
    observations (circles), generated by adding $1.5\%$ Gaussian
    random noise to this surface velocity. The horizontal
    velocity corresponding to the MAP point is shown by the dashed line.
    Right: ``Truth'' sliding coefficient field (solid line) and MAP point
    (dashed line).
  }
  \label{fig:uobs}
\end{figure}
\end{center}

We discuss the performance of the optimization algorithm for the
computation of the MAP point for Problem~2 as defined in
Section~\ref{subsec:app:like}.  On the right in Figure~\ref{fig:uobs},
we show the ``truth'' sliding coefficient field, which is used to
generate the synthetic surface velocity observations. Also shown is
the MAP point, i.e., the solution of \eqref{eq:posterior_V}. In the
upper part of the glacier the MAP point follows the prior mean since
observations are only available in the lower half of the glacier
(i.e., the right part of the domain).

To compute the MAP point, 8 (outer) Newton iterations were
necessary to decrease the nonlinear residual by a factor of
$10^{5}$. In each of these outer Newton iterations, the nonlinear
Stokes equation has to be solved, for which we use an (inner) Newton
method. These inner Newton solves are also terminated after the
residual is decreased by a factor of $10^{5}$, which takes an average of
12 iterations, each amounting to a linearized Stokes solve.
In addition to the nonlinear Stokes solve, each (outer) Newton
iteration requires computation of the gradient and of several
Hessian-vector products. Summing
over all 8 (outer) Newton iterations,
32 conjugate gradient iterations---and thus 32 Hessian-vector
products---are required. In total, the computation of the MAP point amounts to 208
linear(ized) Stokes solves.

\subsection{Low-rank approximation of the prior-preconditioned data misfit
  Hessian}\label{subsec:lowrank} 
\begin{figure}[t]
  \centering\includegraphics[width=0.6\columnwidth]{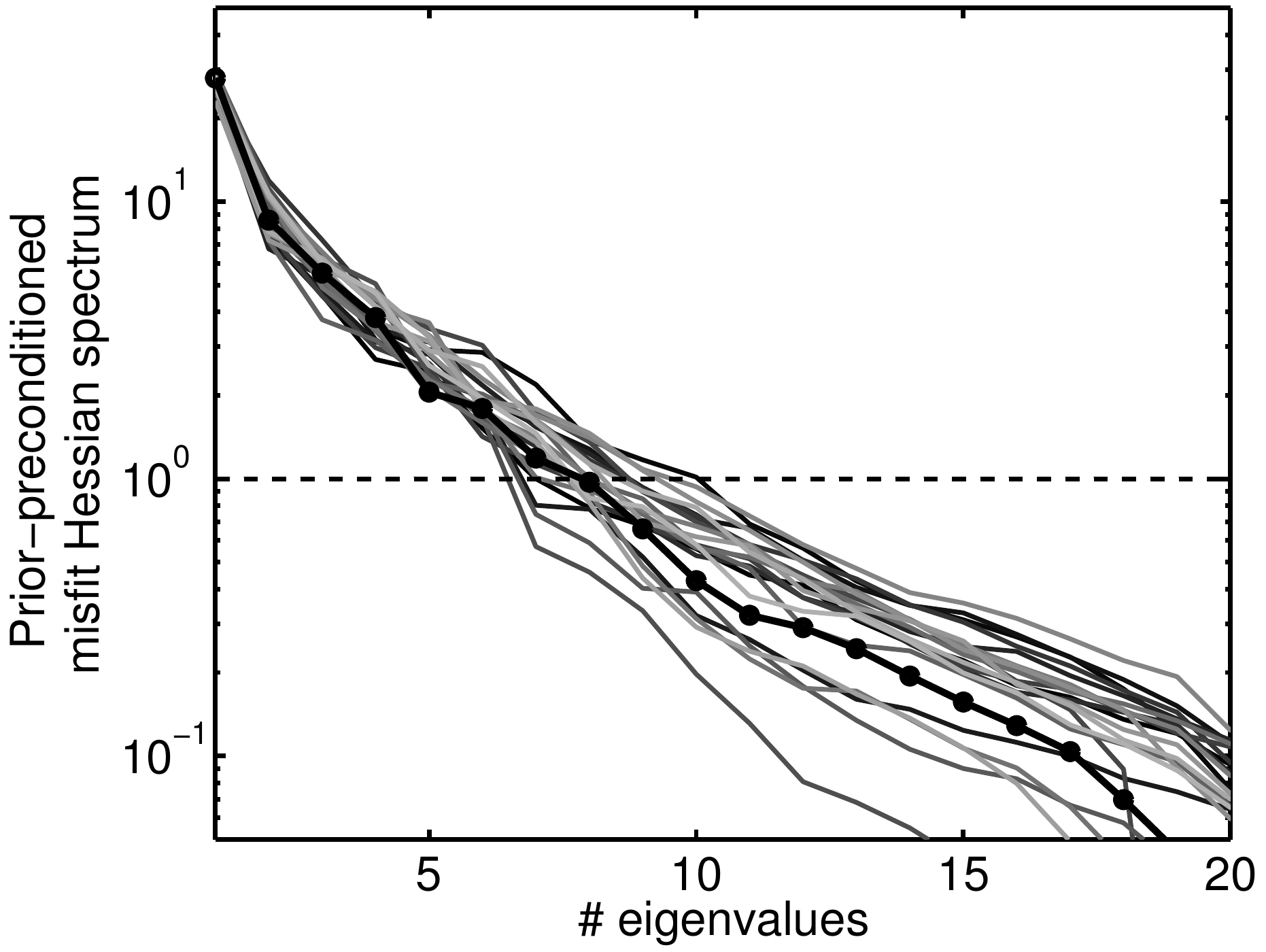}
  \caption{Logarithmic plot of the spectra of prior-preconditioned
    data misfit Hessians computed at the MAP point (black line with dots)
    and at 21 points distributed over the support of the posterior
    (gray lines). %
    The horizontal line for $\lambda = 1$ shows the
    reference value for the truncation of the spectrum of the
    prior-preconditioned data misfit Hessian.}
  \label{fig:eigs}
\end{figure}

The computational feasibility of
Hessian-based sampling for large-scale Bayesian inverse problems
critically relies on low-rank approximations for the data misfit
Hessian. Thus, we study the numerical rank of the prior-preconditioned
data misfit Hessian for various points in the parameter space.
Figure~\ref{fig:eigs} shows a logarithmic plot of the spectra of the
prior-preconditioned data misfit Hessians at the 21 MCMC chain starting
points discussed in Section~\ref{subsec:MCMCcomparison}. Note that all
spectra decay rapidly.  As seen in~\eqref{eq:eigen_error}, an accurate
low-rank approximation of the inverse Hessian can be obtained by
neglecting eigenvalues that are small compared to 1. Thus, retaining
15--20 eigenvectors appears to be sufficient for any point from the
posterior distribution.

In our sampling runs, we thus use $r=20$ eigenvectors for the low-rank
approximation of the prior-preconditioned data misfit Hessian.  We note that the
cost of obtaining this low-rank approximation, measured in the number
of Stokes solves, is $2(r+l)$, where $r+l$ is the number of Lanczos
iterations. Here, $l\ge 0$ iterations are used to ensure the accurate
computation of the most significant eigenvalues/eigenvectors (we use
$l=5$).  We discard any negative eigenvalues to guarantee that the
low-rank approximation is positive semi-definite.

\subsection{Performance of proposed stochastic Newton MCMC method with
  MAP-based Hessian} 
\label{subsec:MCMCcomparison}

In this section, we compare the performance of the proposed SNMAP
method (stochastic Newton MCMC with MAP-based Hessian) with SN (the
original stochastic Newton MCMC method with dynamically-computed
Hessian) and with ISMAP (independence sampler with MAP-based Gaussian)
for both ice flow inverse 
problems introduced in
Section~\ref{subsec:app:like}.
For
each method, 21 MCMC chains are computed using a common set of 21
initial points.  These points are selected from an MCMC chain with
25,000 samples initialized at the MAP point. From this chain, these 21
initial points are chosen to approximately maximize the minimum
pairwise distances between points, so that the resulting set is
distributed quasi-uniformly over the support of the posterior
distribution.  This ensures that the initial points are over-dispersed
with respect to the posterior, which is important for the convergence
diagnostics used to compare the different MCMC methods.

In Table~\ref{tbl:perf}, we summarize convergence diagnostics and MCMC
chain statistics averaged over 21 chains (excluding the MPSRF which is
a multi-chain diagnostic). To compare the different MCMC methods, in
the second column we provide the multivariate potential scale
reduction factor ({MPSRF}) diagnostic~\cite{BrooksGelman98}. This
diagnostic compares averaged properties of individual sample chains
with properties of the pooled sample chain.  When these properties are
similar, we infer that each of the individual sample chains has
converged. The closer the MPSRF is to $1$, the better converged the
individual sample chains are.

It is well known in Monte Carlo methods that the variance in the
estimate decays as $1/N$ when averaging over $N$
i.i.d.\ samples.
  However, MCMC samples are 
not independent, and in general we observe that averaging
over $N$ samples from an MCMC chain reduces the variance in the
estimate by a factor of only $\tau/N$, where $\tau > 1$ is the
integrated autocorrelation time ({IAT})\cite{RobertCasella04},
given by
\begin{equation}
\tau = 1 + 2 \sum_{s=1}^\infty \rho (s).
\label{integratedautocorrelation}
\end{equation}
Here, $\rho(s)$ is the usual autocorrelation function for a lag $s>0$.
In practice, $\rho(s)$ is noisy when computed from a finite
length sample chain, and thus we estimate $\tau$ by the maximum value
obtained by truncating the summation in
\eqref{integratedautocorrelation}. The autocorrelation is defined for
a scalar quantity, and we report in column three the IAT corresponding
to the sliding coefficient field at the point $x = 3450$.
In the fourth column, we report the
effective sample size (ESS) defined as $N/\tau$,
the number of independent samples that would be required for the same
variance reduction as obtained from the MCMC chain.

The fifth column shows the mean squared jump distance ({MSJ}),
which provides an indication of how well the MCMC chain is mixing.
This metric is defined for a
single MCMC chain with samples $\params_0, \ldots, \params_N$ as
\begin{equation}
\mathrm{MSJ} := \frac 1N \sum_{k=0}^{N-1} \| \params_{k+1} -
\params_{k} \|_{\M}^2.
\end{equation}
In general, a larger mean square jump
distance indicates faster mixing of the MCMC chain, and tends to
result in better chain convergence.

Finally, we address the question of greatest
interest with regard to computational efficiency: ``Given an MCMC
algorithm, how much computational work is required to obtain
an independent sample?''.  Column seven reports the total number of
linearized Stokes solves required %
to obtain a single independent sample,
and column eight reports the total wall-clock time for these solves.
\begin{table}[t]
    \caption{Multivariate potential scale reduction factor ({\bf
        MPSRF}), integrated autocorrelation time ({\bf IAT}),
      effective sample sample size ({\bf ESS}), mean squared jump
      distance ({\bf MSJ}), acceptance rate ({\bf AR}), number of
      (linearized) Stokes solves per independent sample ({\bf SPIS}),
      and the average wallclock time per independent sample ({\bf
        TPIS}). We compare the performance obtained with the
      independence sampler with MAP-based Gaussian ({ISMAP}), the
      stochastic Newton MCMC method with MAP-based Hessian ({SNMAP}),
      and the stochastic Newton MCMC method with dynamic Hessian
      ({SN}) for two problems with different noise levels (e.g.,
      $\bar\sigma_{\scriptsize{\text{noise}}} = 62$ and
      $\bar{\bar\sigma}_{\scriptsize{\text{noise}}} = 10$ for Problem
      1, and $\bar\sigma_{\scriptsize{\text{noise}}} = 18$ and
      $\bar{\bar\sigma}_{\scriptsize{\text{noise}}} = 3$ for Problem
      2). We use 21 MCMC chains, each with 25,000 samples, hence the
      total number of samples is 525,000. The dimension of the
      discretized basal sliding coefficient field, i.e., the number of
      parameters, is 139.}
  \begin{center}
    \begin{tabular}{|l|c|c|c|c|c|c|c|}
      \hline\hline
          {} & {\bf MPSRF} & {\bf IAT} & {\bf ESS} & {\bf
            MSJ} & {\bf AR ($\%$)} & {\bf SPIS} & {\bf TPIS (s)}\\
          \hline\hline
          \multicolumn{8}{|c|}{{\bf Problem 1}}\\
          \hline
          {ISMAP} & 1.210 & 253 & 2075  & 1456 & 41 & 2783  & 139\\
          {SNMAP}  & 1.001 & 6 & 84004 & 1390 & 40 & 72     & 4  \\
          {SN} & 1.073 & 125 & 4032  & 565  & 17 & 1375 & 69\\
          \hline
          \multicolumn{8}{|c|}{{\bf Problem 2}}\\
          \hline
          {ISMAP} & 1.507 & 435 & 1207 & 280 & 9  & 4350 &  218\\
          {SNMAP}  & 1.045 & 80 & 6563 & 190  & 6  & 960  & 48 \\
          {SN} & 1.348 & 600 & 875  & 64   & 2  & 8400 & 420 \\
          \hline
    \end{tabular}
    \label{tbl:perf}
  \end{center}
\end{table}

We summarize the following observations from Table~\ref{tbl:perf}:
\renewcommand{\labelenumi}{(\alph{enumi})}
\begin{itemize}
\item The number of independent samples is about one order of
  magnitude larger for Problem~1 than for Problem~2, suggesting that
  the posterior distribution for Problem~2 is %
  more difficult to sample.

\item SNMAP leads to the best MPSRF values for both
  problems, suggesting the fastest convergence with respect to the
  number of samples. As a consequence, the largest effective sample
  size is achieved using SNMAP.
  Note that this holds even though ISMAP yields larger acceptance rates and mean
  squared jump distances.

\item SNMAP also requires the smallest number of forward
  solves per independent sample. For Problem~1, SNMAP is more
  efficient than SN by a factor of about 20, and than ISMAP by a factor
  of almost 40. For Problem 2,
  SNMAP is more efficient by factors of about 10 and 5
  than SN and ISMAP, respectively.

\item Surprisingly, SN performs worse than SNMAP, even with respect to
  the number of samples. This is despite the fact that it uses a
  better local approximation of the posterior. We attribute this to
  the mismatch in the local Hessians at different points, which
  increases the asymmetry between the forward and backward proposals
  $q(\params_k,\proposal)$ and $q(\proposal,\params_k)$, thus
  increasing the variability of the acceptance probability
  $\alpha_{k}(\proposal)$ (see Algorithm~\ref{algorithm:M-H}).

\end{itemize}
We have also applied Delayed Rejection Adaptive Metropolis (DRAM)
sampling~\cite{HaarioLaineMiraveteEtAl06} to explore the posterior
distribution. We found it to be far from convergence after 1,000,000
samples. We attribute this to the high
dimensional parameter space and the lack of information about
the problem structure in the sampling process. In
the next section, we focus on visualization and interpretation of the
posterior distribution.

\section{Analysis and interpretation of the solution of the
  Bayesian inverse problem}
\label{sec:uqinterpr}

Visualization and interpretation for high-dimensional posterior distributions
is a difficult task.
In this section, we highlight techniques motivated by the structure
of the Bayesian inverse problem to guide our analysis.
First, in Section~\ref{sec:point_marginals}, we present visualizations of the posterior
in the physical coordinate basis, which provides intuition about the
solution at particular points or regions of the domain.
Then, in Section \ref{sec:eigenvector_postcov}, we shift our
perspective to eigenvectors of the posterior covariance
(approximated using the Hessian at the MAP point), which can be
classified into groups according to their contributions from the observation data and the prior.
The qualitative features of each group
provide insight into the posterior distribution.
Finally, in Section~\ref{sec:eigenvector_marginals}, we visualize one-
and two-dimensional marginal distributions of the full posterior
distribution with respect to these eigenvectors.

The results discussed in this section are for Problem~2 as defined in
Section~\ref{subsec:app:like}, i.e., the
problem with smaller data noise.  The approximation of the posterior
pdf is based on samples generated by the SNMAP method, and kernel
density estimation is used to visualize the one- and two-dimensional
marginal probability density functions.

\subsection{Point marginals and samples from the posterior}
\label{sec:point_marginals}

\begin{center}
\begin{figure}[t!]
  \includegraphics[width=0.49\columnwidth]{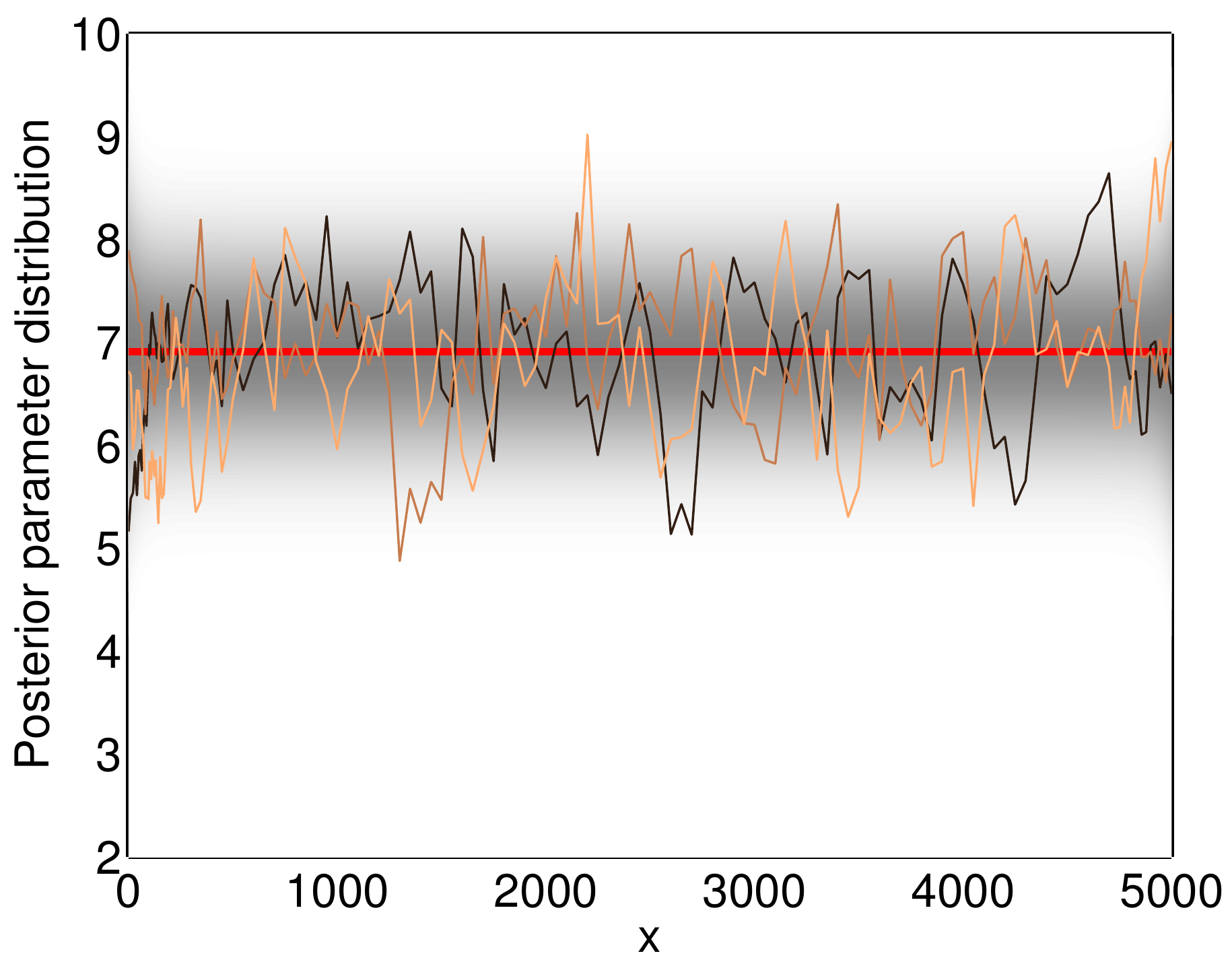}
  \includegraphics[width=0.49\columnwidth]{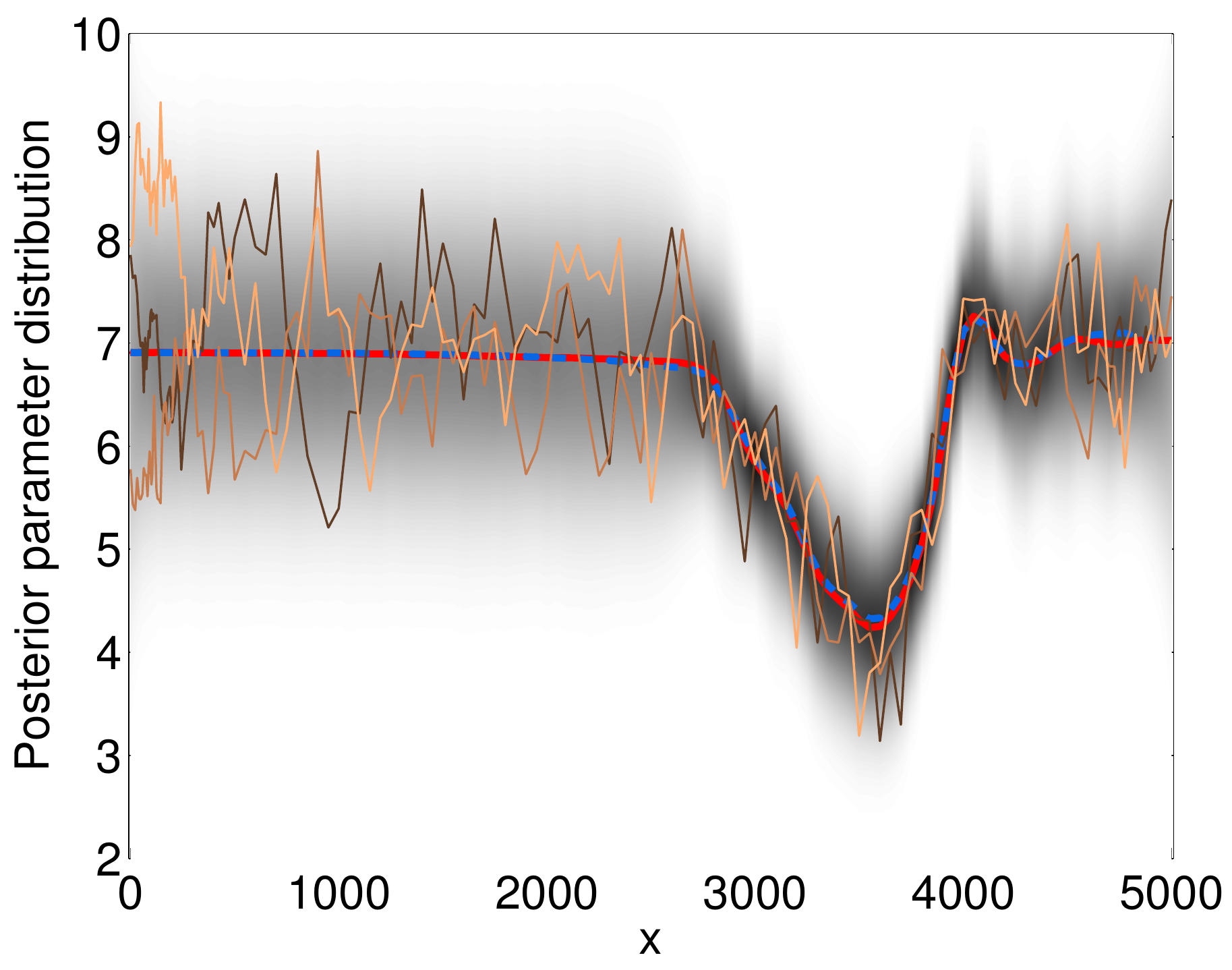}
  \caption{Shown in gray scale are the one-dimensional point marginals of the prior
    (left) and the posterior (right) probability distributions,
    with higher probability density indicated by darker shading.
    Point marginals are computed and plotted independently along a
    vertical line at each point, where the gray shaded area corresponds to a
    95\% confidence interval.
    To give an indication of spatial correlation, samples from the
    prior and the posterior are shown (in different shades of brown).
    Also shown are the prior and posterior mean (in red), and the MAP
    point of the posterior (in blue). We recall that the dimension of
    the discretized basal sliding coefficient field, i.e., the number
    of parameters, is 139.  \label{fig:prior_post_gradient}}
\end{figure}
\end{center}

In Figure~\ref{fig:prior_post_gradient}, we present (one-dimensional)
marginals of the prior and posterior distributions with respect
to physical points in the domain. We refer to these marginals as point
marginals.
The probability density for each point marginal is visualized in gray scale
along a vertical column at each point, with higher probability density indicated
by darker shading.
Because each point marginal is computed independently, the point
marginal density values at neighboring points are not necessarily
related, and thus any spatial correlation structure present in the
distribution is neglected by this visualization.
For this reason, we overlay a few samples from each
distribution to provide some indication of the spatial correlation structure.

This visualization provides some useful observations for our problem.
In the unobserved part of the domain (the upper part of the glacier),
the point marginals of the posterior are similar to those of the
prior; our beliefs about the basal sliding coefficient field in this region are
unchanged from the prior.
On the contrary, in the region where observation data are available, we
find the variance to be decreased significantly (i.e., we are more
certain about the sliding coefficient field in this region),
and in some regions most of the probability mass is shifted in the
posterior compared to the prior; the evidence from the observation data
has overwhelmed our prior beliefs in this region.
Finally, while spatial correlation structure is difficult to
infer from the limited number of overlayed samples, note %
that the average width of the variations %
appears unchanged from the prior to the posterior in both, the parts
of the glacier with and without observations.  We interpret this as
insufficient observational evidence to update our beliefs about the
width of spatial variations.

\subsection{Classification of posterior covariance eigenvectors}
\label{sec:eigenvector_postcov}

In this section, we classify the eigenvectors of the posterior
covariance into groups according to their contributions from the
observation data and prior, and subsequently use this classification
to gain insight into the posterior distribution.
While it is common to order
eigenvectors by ascending or descending eigenvalues, this choice is
poorly adapted to our purposes since it unpredictably interleaves
data-influenced eigenvectors with prior-influenced eigenvectors.
We therefore propose a general technique for sorting eigenvectors
that groups them naturally.

\begin{figure}[t]
\begin{center}
  \begin{tikzpicture}
    \node (img3) at (-0.12\columnwidth,
    -0.51\columnwidth){\includegraphics[width=0.48\columnwidth]{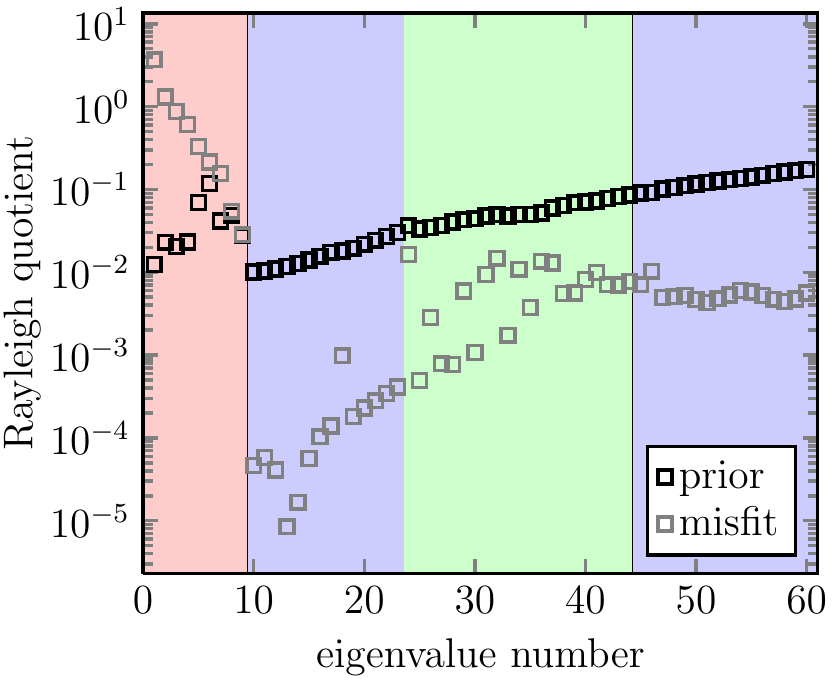}};
    \node (img1) at (0.36\columnwidth,
    -0.516\columnwidth){\includegraphics[width=0.47\columnwidth]{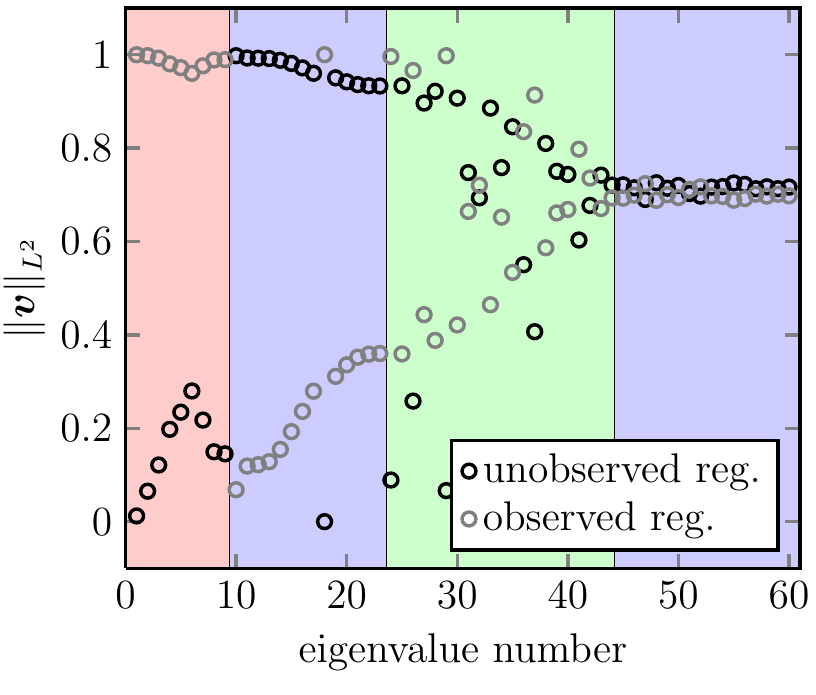}};
  \end{tikzpicture}
  \caption{Left: Semilogarithmic plot of the Rayleigh quotients of the
    data misfit Hessian and of the inverse prior covariance as defined
    in~\eqref{eq:r_coeffs}. Right: Norm of the eigenvectors in the lower
    and upper parts of the glacier, i.e., in the region with and without
    observations, respectively.\label{fig:rayleigh}}
\end{center}
\end{figure}

To characterize the influence of the observations and prior on the
eigenvectors, consider the Rayleigh quotients of the data misfit Hessian
and of the inverse of the prior, i.e.,
\begin{align}\label{eq:r_coeffs}
  r_{\!m}^i = \frac{\mip{\bs v_i}{\H_{\text{misfit}} \bs v_i}}{\mip{\bs v_i}{\bs v_i}}, \quad
  r_{\!p}^i = \frac{\mip{\bs v_i}{\matrix{\Gamma}_{\text{prior}}^{-1}\bs
      v_i}}{\mip{\bs v_i}{\bs v_i}},
\end{align}
for $i = 1,\ldots,n$, where $\bs v_i$ is the $i$-th eigenvector of the
inverse posterior covariance (approximated by the Hessian at the MAP point).
Because the eigenvalue $\lambda^i$ associated with $\bs v_i$ is simply
the sum of $r^i_{\!m}$ and $r^i_{\!p}$, these Rayleigh quotients
quantify the contributions from the observation data
and prior.
We then order the eigenvectors according to
the difference of the squared Rayleigh coefficients
$d^i:=(r^i_{\!m})^2-(r^i_{\!p})^2$. Large positive values of $d^i$
correspond to eigenvectors that are most informed
by the data, whereas large negative values correspond to directions most informed
by the prior.
We note that there are several reasonable choices
for $d^i$; we find that our choice best groups
eigenvectors with similar qualitative features.
The sorted Rayleigh quotients for the data misfit Hessian and for
the inverse of the prior are presented in the left plot in
Figure~\ref{fig:rayleigh}.  A selection of these eigenvectors is shown
in Figure~\ref{fig:ordemodesh}.

Next, we study the qualitative features of these
eigenvectors. %
Since the lower half
of the glacier contains observation points and the upper half
does not, we
can also characterize these eigenvectors
by determining whether the eigenvector is concentrated primarily
in one half of the glacier.
The right plot in Figure~\ref{fig:rayleigh} studies these concentrations
in each half of the domain using the
corresponding $L^2$-norms.  We can distinguish four groups of eigenvectors,
highlighted by different colors in
Figure~\ref{fig:rayleigh}, which we discuss next.

\setlength{\tabcolsep}{1pt} \def \pos {0.32\columnwidth}
\begin{center}
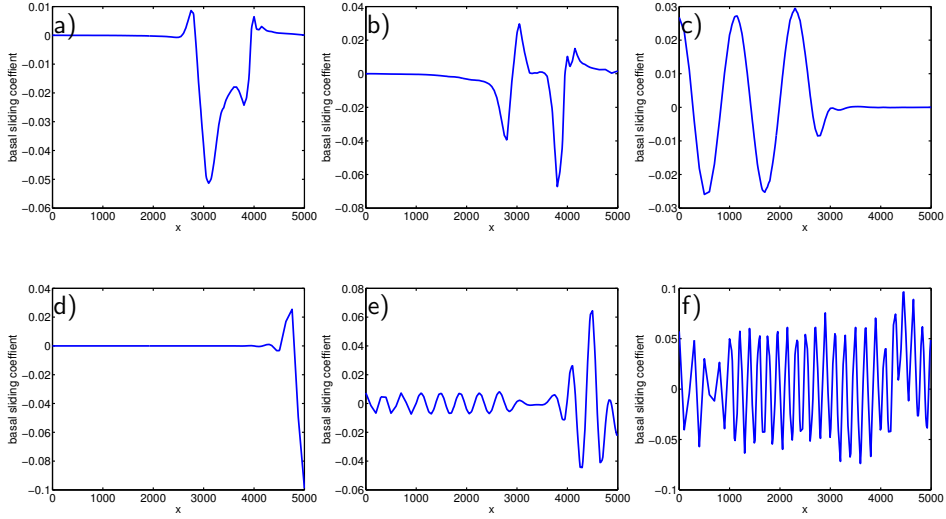
\begin{figure*}
  \begin{tikzpicture}
    \node (11) at (0*\pos, 1.9*\pos) {
      \includegraphics[width=0.32\columnwidth]{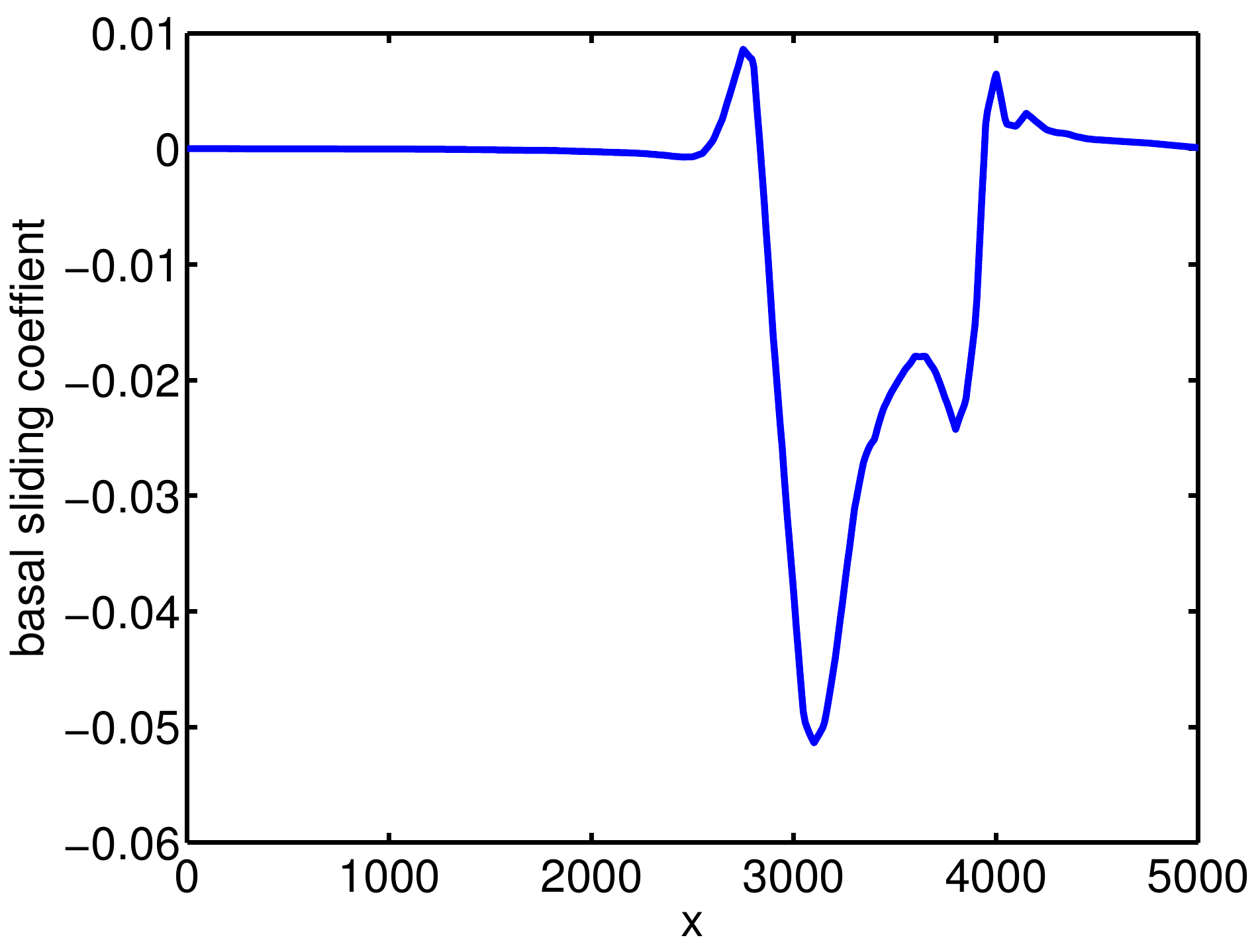}};
    \node (12) at (1*\pos, 1.9*\pos) {
      \includegraphics[width=0.32\columnwidth]{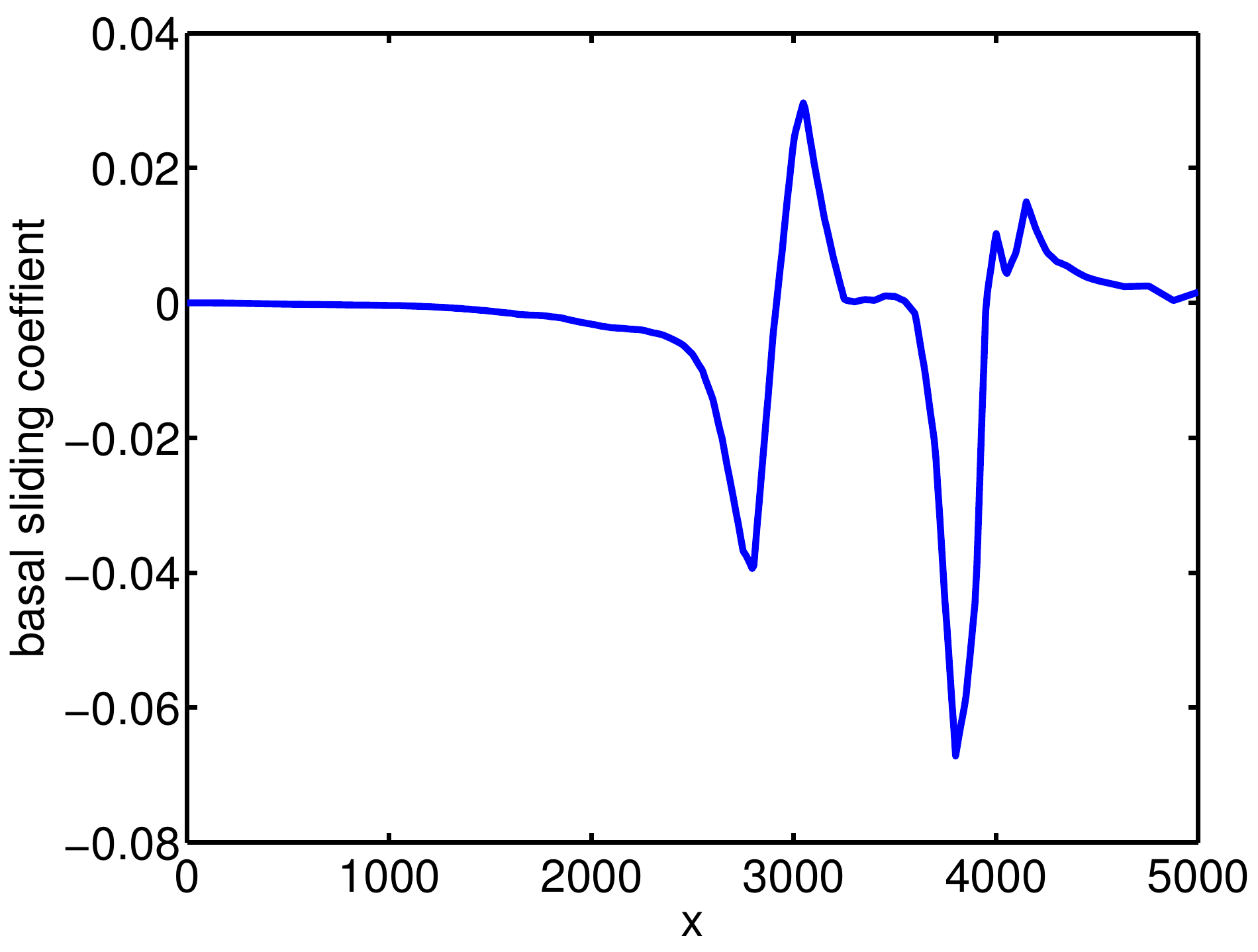}};
    \node (13) at (2*\pos, 1.9*\pos) {
      \includegraphics[width=0.32\columnwidth]{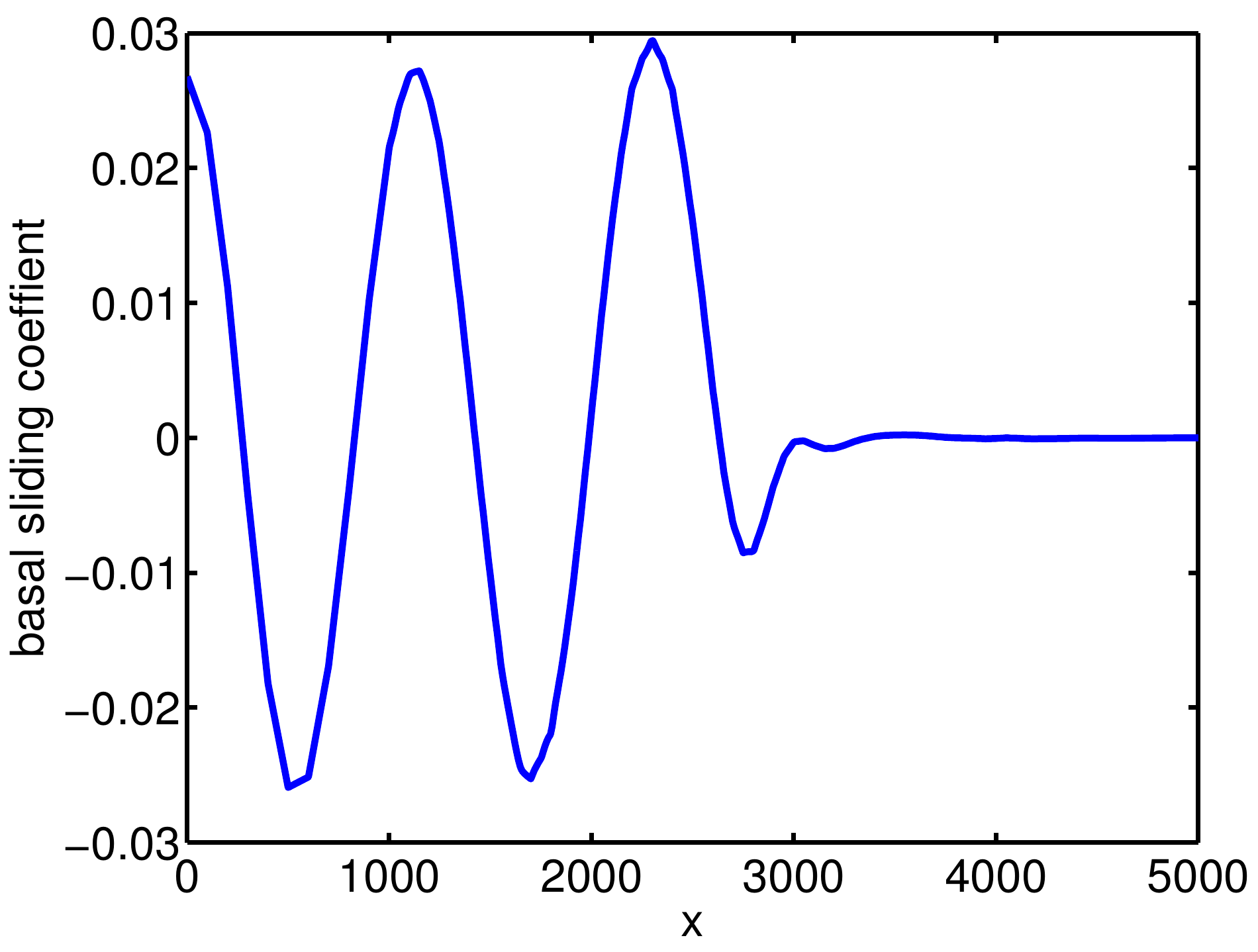}};
    \node (21) at (0*\pos, 1*\pos) {
      \includegraphics[width=0.32\columnwidth]{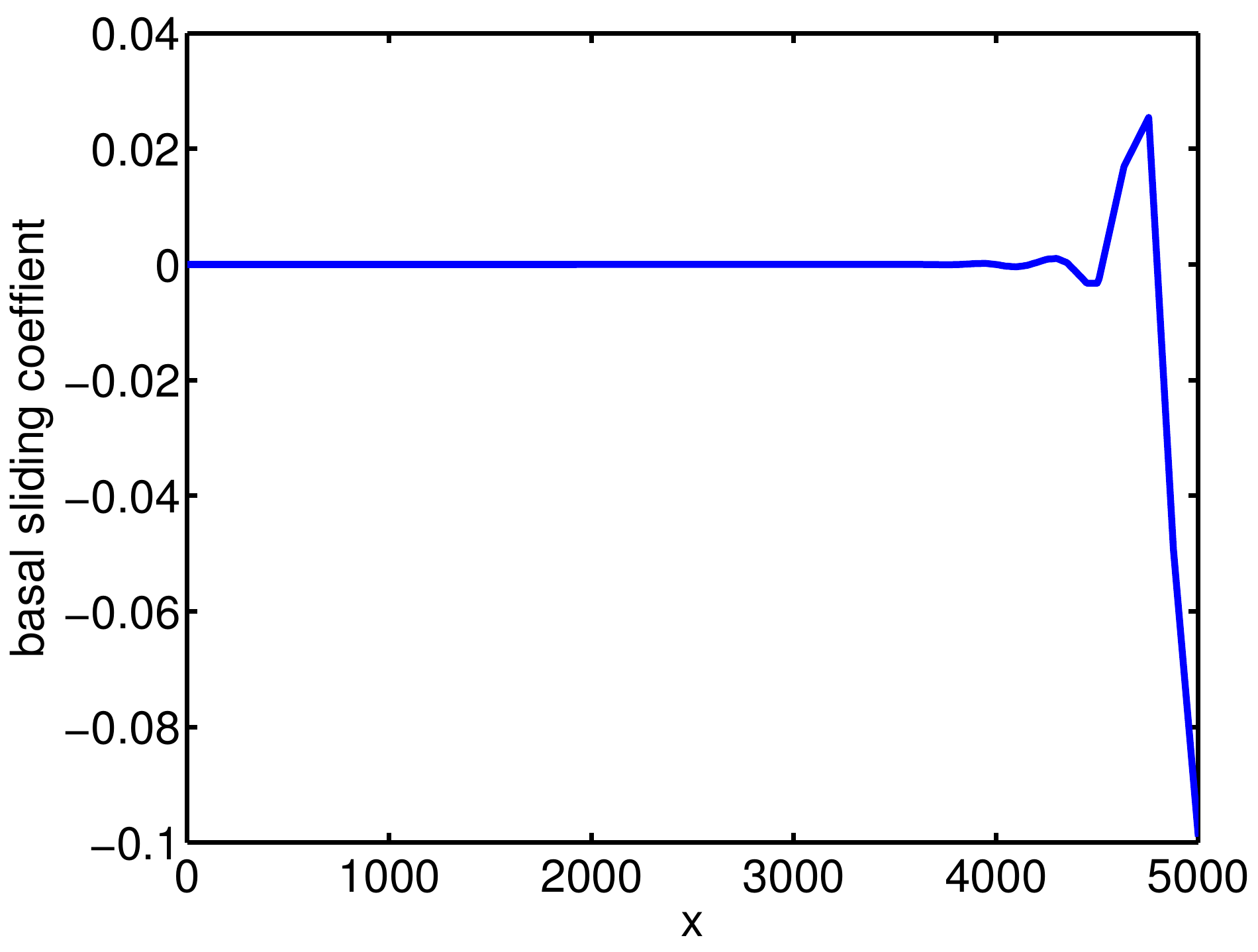}};
    \node (22) at (1*\pos, 1*\pos) {
      \includegraphics[width=0.32\columnwidth]{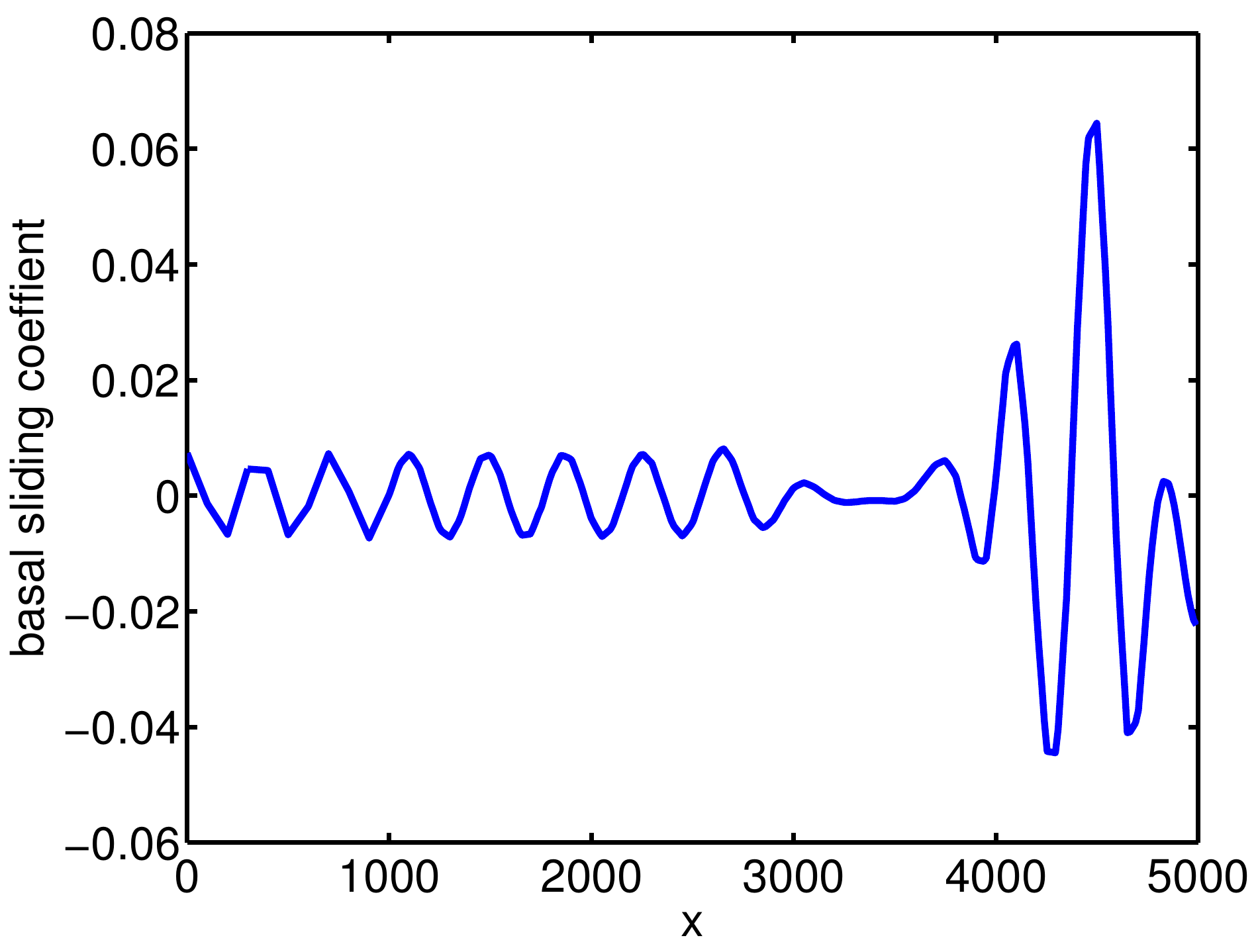}};
    \node (23) at (2*\pos, 1*\pos) {
      \includegraphics[width=0.32\columnwidth]{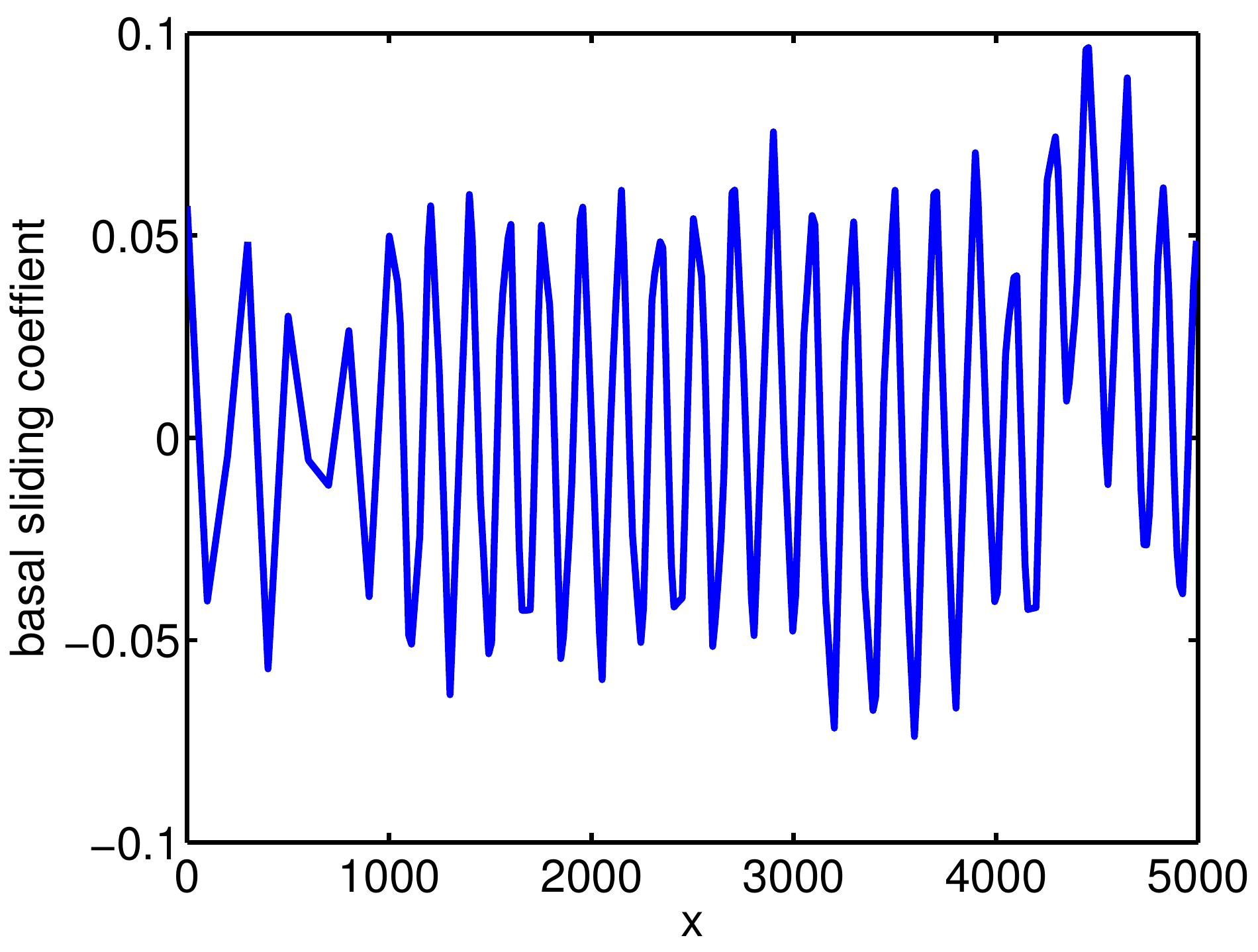}};
    \node at (0*\pos-0.31*\pos, 1.9*\pos+0.29*\pos) {\sf a)};
    \node at (1*\pos-0.31*\pos, 1.9*\pos+0.29*\pos) {\sf b)};
    \node at (2*\pos-0.31*\pos, 1.9*\pos+0.29*\pos) {\sf c)};
    \node at (0*\pos-0.31*\pos, 1*\pos+0.29*\pos) {\sf d)};
    \node at (1*\pos-0.31*\pos, 1*\pos+0.29*\pos) {\sf e)};
    \node at (2*\pos-0.31*\pos, 1*\pos+0.29*\pos) {\sf f)};
  \end{tikzpicture}
  \caption{Eigenvectors of the Hessian corresponding to the 1st, 3rd,
    14th, 18th, 26th and 55th eigenvalues %
    are shown in figures a--f, respectively. Note that different
    eigenvectors are concentrated in different parts of the domain and
    that eigenvectors corresponding to smaller eigenvalues are more
    oscillatory.
    \label{fig:ordemodesh}
  }
\end{figure*}
\end{center}

\subsubsection*{Data-informed eigenvectors}
The first group (shown in red in Figure~\ref{fig:rayleigh})
contains eigenvectors for which $d^i$ is positive, i.e., the information from the data
dominates the information from prior.  In the direction of these eigenvectors, the
variance in the posterior is significantly reduced due to the
observations (recall that the variance is
$1/\lambda^i = 1/(r^i_{\!m} + r^i_{\!p})$~), 
and hence we say they have been informed by the data.

The eigenvectors in this group are primarily concentrated in the lower
half of the glacier where we have observations (see the right plot in
Figure~\ref{fig:rayleigh}). They are relatively smooth (since
$r^i_{\!p}$ is not large), and qualitatively resemble the first nine
Fourier modes in this region (see plots (a) and (b) in
Figure~\ref{fig:ordemodesh} for eigenvectors~1 and~3).
This last observation is powerful since it provides
confidence that features of the MAP point %
that lie in the span of
these first nine Fourier modes are indeed features of the
true basal sliding coefficient field.

\subsubsection*{Shadowed eigenvectors}

The next group contains eigenvectors for which the original prior
variance was large, and yet the observations provide
little information (they are not illuminated by the data, and thus
``shadowed'').
These eigenvectors are characterized by large ratios
$r^i_{\!p}/r^i_{\!m}$, and the posterior is easy to characterize in
these directions; it is similar to the prior.
This group as well as the prior-tail group discussed below are shown in blue in
Figure~\ref{fig:rayleigh}. Both groups are well
characterized by the prior distribution, although for different reasons.

Shadowed eigenvectors generally concentrate in regions where the
parameter-to-observable map is insensitive to the parameter.
In our problem, the upper part of the glacier is far away from
observation points, and the basal sliding coefficient field at a point only
has significant influence on the ice velocity in a neighborhood of that point.
Thus, the parameter-to-observable map is insensitive to the sliding
coefficient in the upper part of the glacier.
In Figure~\ref{fig:rayleigh} we can see that indeed most
of these eigenvectors are concentrated in the upper part of the glacier, and
again resemble Fourier modes in the upper half of the glacier
(see Figure~\ref{fig:ordemodesh}c).

Parameter-to-observable map insensitivity also occurs
at the very bottom edge of the glacier.  Even though observations
are available here, the flow in this region
is determined  primarily by the glacial boundary, preventing the basal
sliding coefficient field from significantly influencing the surface
velocity in this region.
Exactly one shadowed eigenvector corresponds to this region, shown in
Figure~\ref{fig:ordemodesh}d.

\subsubsection*{Mixed eigenvectors}
The third group contains eigenvectors for which
the observations and the prior both have a significant
influence.
In general it is not clear how this interaction will affect the
posterior distribution, and as such it is perhaps too optimistic to
make predictions based on this analysis
and we defer this discussion to Section~\ref{sec:eigenvector_marginals}.
Note that these eigenvectors seem to be generally characterized by a
mixture of medium frequency Fourier-like modes on the upper and lower
half of the glacier, which is why we refer to them as ``mixed'' eigenvectors.
One eigenvector from this group is shown in Figure~\ref{fig:ordemodesh}e.

\subsubsection*{Prior-tail eigenvectors}
The remaining posterior eigenvectors represent directions in
which the prior is very certain (i.e., the prior variance
$1/r^i_{\!p}$ is small), and for which the observations do not provide
sufficient evidence to either contradict or reinforce this assertion
(i.e., as in the shadowed eigenvectors, the ratio
$r^i_{\!p}/r^i_{\!m}$ is large).
In the continuous inverse problem, this final group contains an infinite number
of eigenvectors, each behaving very similar to their prior counterparts, and
we therefore refer to this as the ``prior-tail''.
One eigenvector  from this prior-tail group,
which qualitatively resembles a high frequency Fourier mode,
is shown in Figure~\ref{fig:ordemodesh}f.

\subsection{Marginals in the eigenvector directions}
\label{sec:eigenvector_marginals}

While the analysis of the previous section provides some
insight into the posterior distribution, it has two important
limitations.
First, this analysis is predicated on the assumption that the
posterior is completely characterized by its mean and covariance, and
as such, any non-Gaussian behavior of the posterior is obscured.
Second, the analysis makes use of the posterior covariance approximated
at the MAP point, which may not reflect the behavior of the
posterior away from this point.
In this section, we make use of the insights gleaned from the above
analysis, but return our focus to %
the full posterior distribution.

\begin{figure}[t]
\begin{center}
  \begin{tikzpicture}
    \node (img1) {\includegraphics[width=0.65\columnwidth]{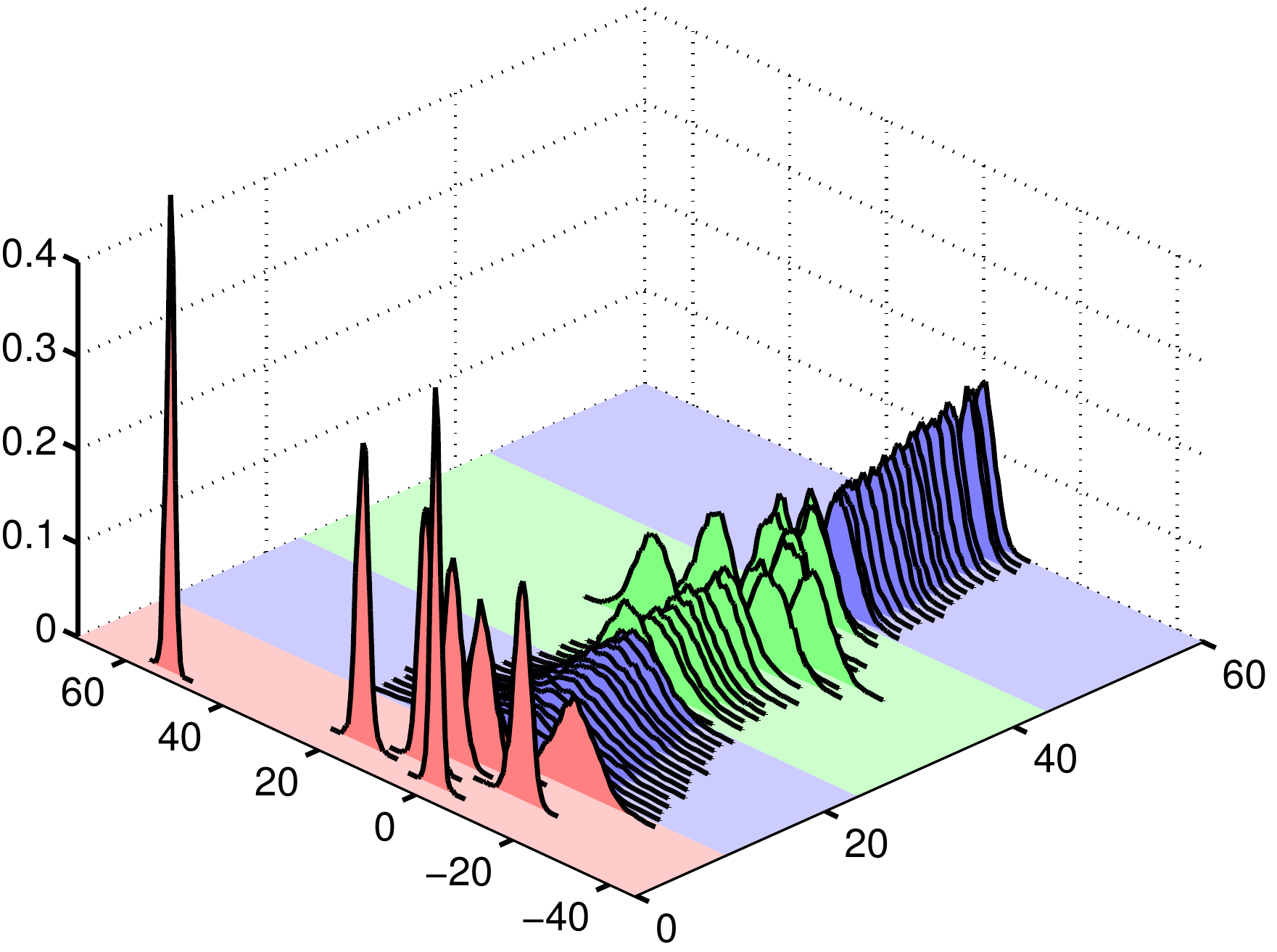}};
    \node[rotate=25] at (2.5,-2.5) {eigenvalue number};
    \node (img2) at (0.36\columnwidth,
    0.2\columnwidth){\includegraphics[width=0.47\columnwidth]{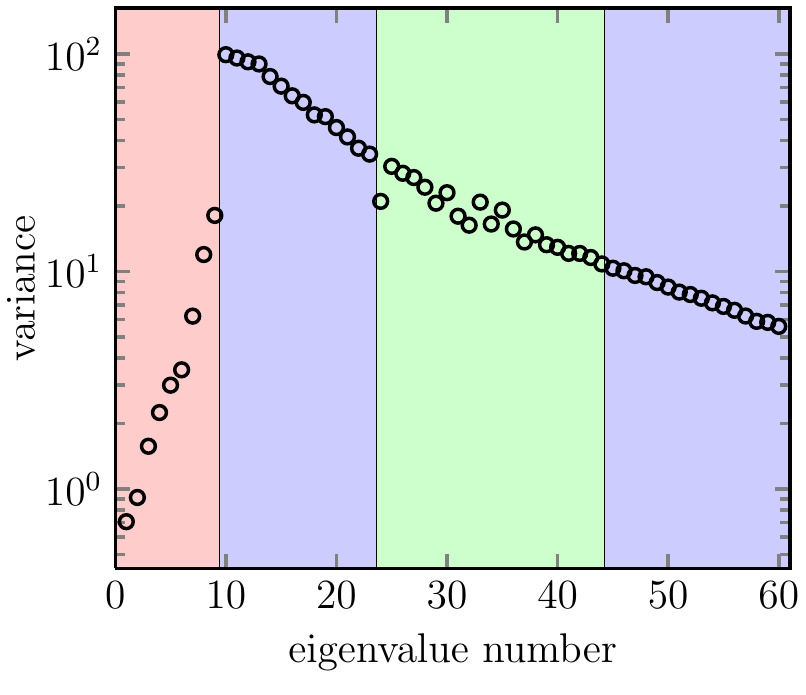}};
    \node[rotate=-25] at (-2.5,-2.5) {parameter};
  \end{tikzpicture}
  \caption{Marginals of the posterior distribution with respect to eigenvectors
    of the covariance (approximated using the Hessian at the MAP point).
    Shown are
    the marginals (left) and the corresponding sample variances (right).
    The eigenvectors are sorted with respect to qualitative features
    indicated by different background colors as described in the text.
  }
  \label{fig:stack_pdf_prior}
\end{center}
\end{figure}

In Figure~\ref{fig:stack_pdf_prior}, we show the one-dimensional
marginals and sample variances of the posterior distribution, with respect to
eigenvectors of the posterior covariance using colors corresponding
to the eigenvector groups discussed in
Section~\ref{sec:eigenvector_postcov}. Many features of these
marginals are already anticipated: the data-informed eigenvectors
(in red) have small variance and are most shifted with respect to the
prior distribution.  The shadowed eigenvectors (first blue group)
have the largest variances and the prior-tail eigenvectors (second
blue group) have small variance and are essentially unchanged from the
prior. To emphasize the departure of the posterior from the prior, all
marginals are plotted with respect to the prior mean. Any shift of the
marginal away from zero is due to observations.

Despite the nonlinearity of the parameter-to-observable map, we find
that the posterior marginals all appear to be near-Gaussian.
Since the noise and prior models are both Gaussian, it is reasonable
to expect gaussianity of the data-informed directions in the
small-noise limit (the parameter-to-observable map is smooth and thus
nearly linear over a narrow range), and also in the directions where
the prior is most influential, as the data does not
update the prior distribution in these directions.
We therefore anticipate that the most non-Gaussian behavior
occurs in the mixed eigenvector directions (in green),
as these are the directions with the largest variance (so that the
parameter-to-observable map can deviate from a linear approximation)
that are significantly influenced by the data.

Figure~\ref{fig:2dmarginals_hesslike},
depicts one- and two-dimensional marginals of the posterior distribution in selected
eigenvector directions
together with the Gaussian approximation of
the posterior distribution at the MAP point. As in
Figure~\ref{fig:stack_pdf_prior}, these marginals are plotted with
respect to the prior mean.
In all directions except for the first (the most data-informed eigenvector),
we observe that the posterior marginal is close to %
the Gaussian approximation at the MAP point even in the mixed eigenvector
direction ($v_{26}$). 
In the direction of the first eigenvector, there is a
clear shift in the marginal mean of the posterior distribution and its Gaussian
approximation at the MAP point. Nevertheless,
the corresponding posterior marginal looks Gaussian.
To give a possible explanation for this behavior, consider a
two-dimensional pdf with banana-shaped contours for which
the MAP point is located along the banana ridge,
but the mean is located at the banana's center of mass, in a region
that itself may have low probability density.  One-dimensional
marginals of such a pdf are likely to have a similar discrepancy
between the MAP point and the mean.
Although with respect to the other eigenvectors, the marginals of
the posterior and the Gaussian approximation at the MAP point are
close, this does not necessarily imply that the posterior is Gaussian.

\begin{center}
  \setlength{\tabcolsep}{1pt}
  \def \pos  {0.25\columnwidth}
  \def \posy {0.205\columnwidth}
  \begin{figure}
    \begin{tikzpicture} [node distance=0cm,auto]

    \node (L21) at (0*\pos, 2*\posy)
         {\includegraphics[width=0.24\columnwidth]{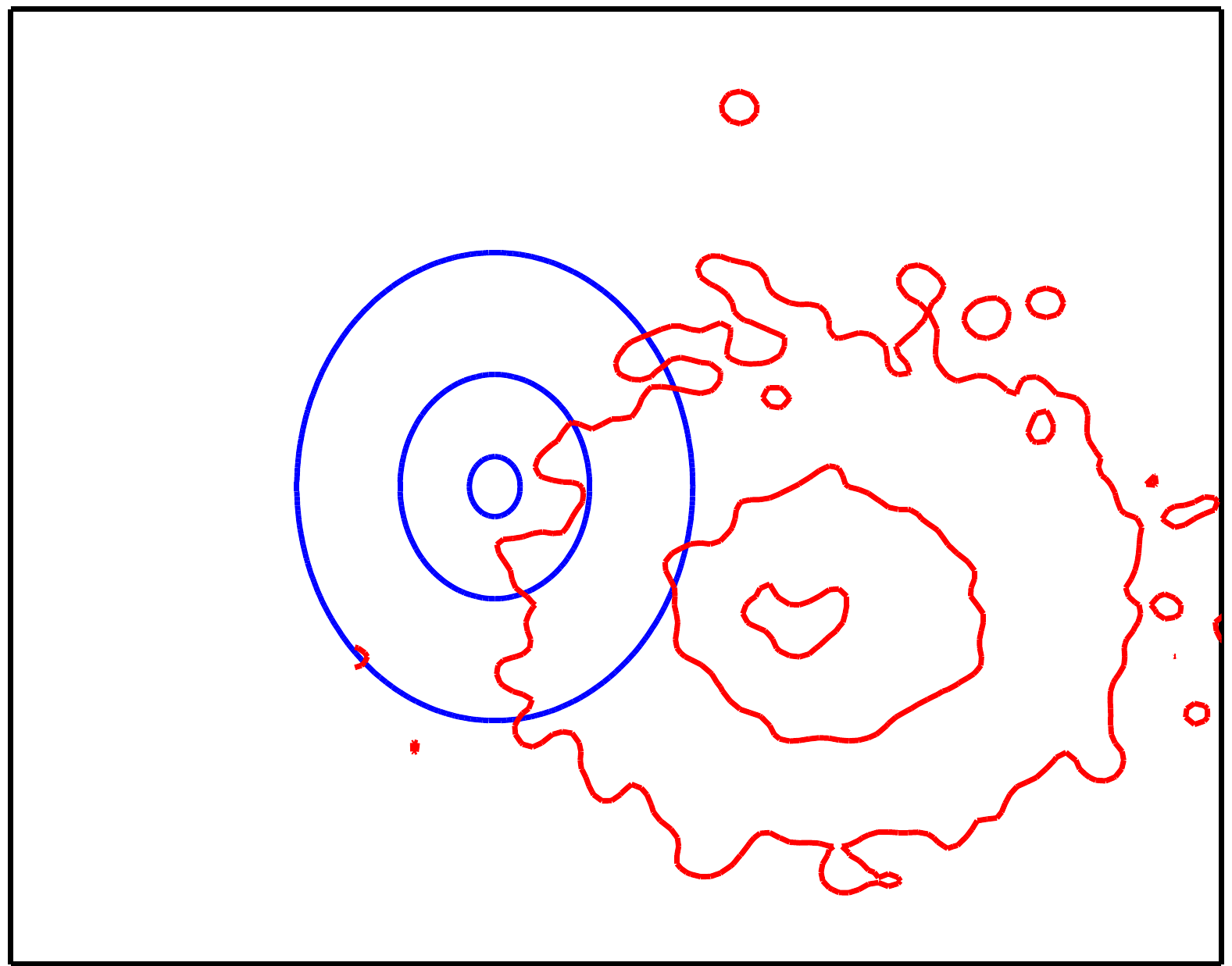}};
    \node (L31) at (0*\pos, 1*\posy)
         {\includegraphics[width=0.24\columnwidth]{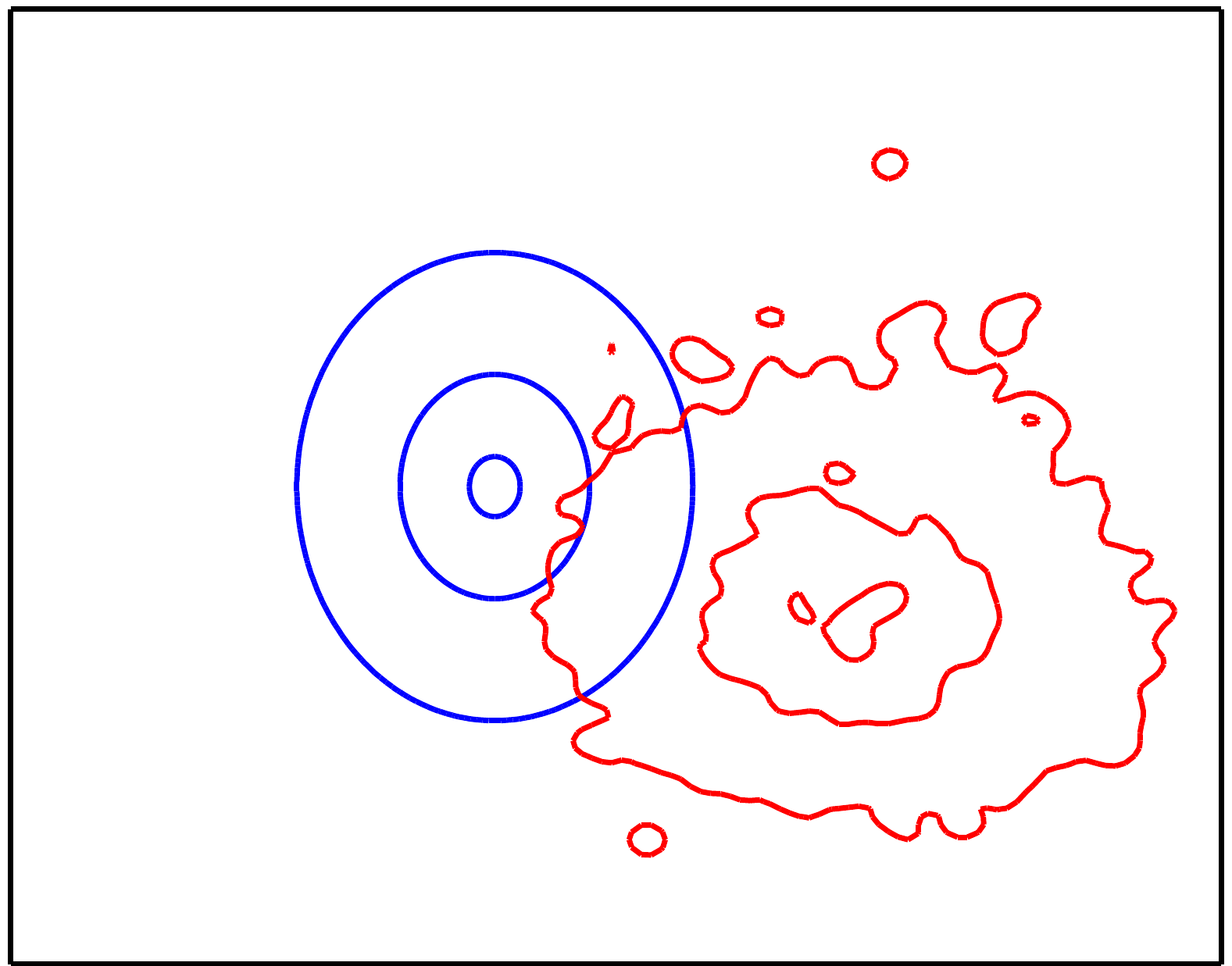}};
    \node (L32) at (1*\pos, 1*\posy)
         {\includegraphics[width=0.24\columnwidth]{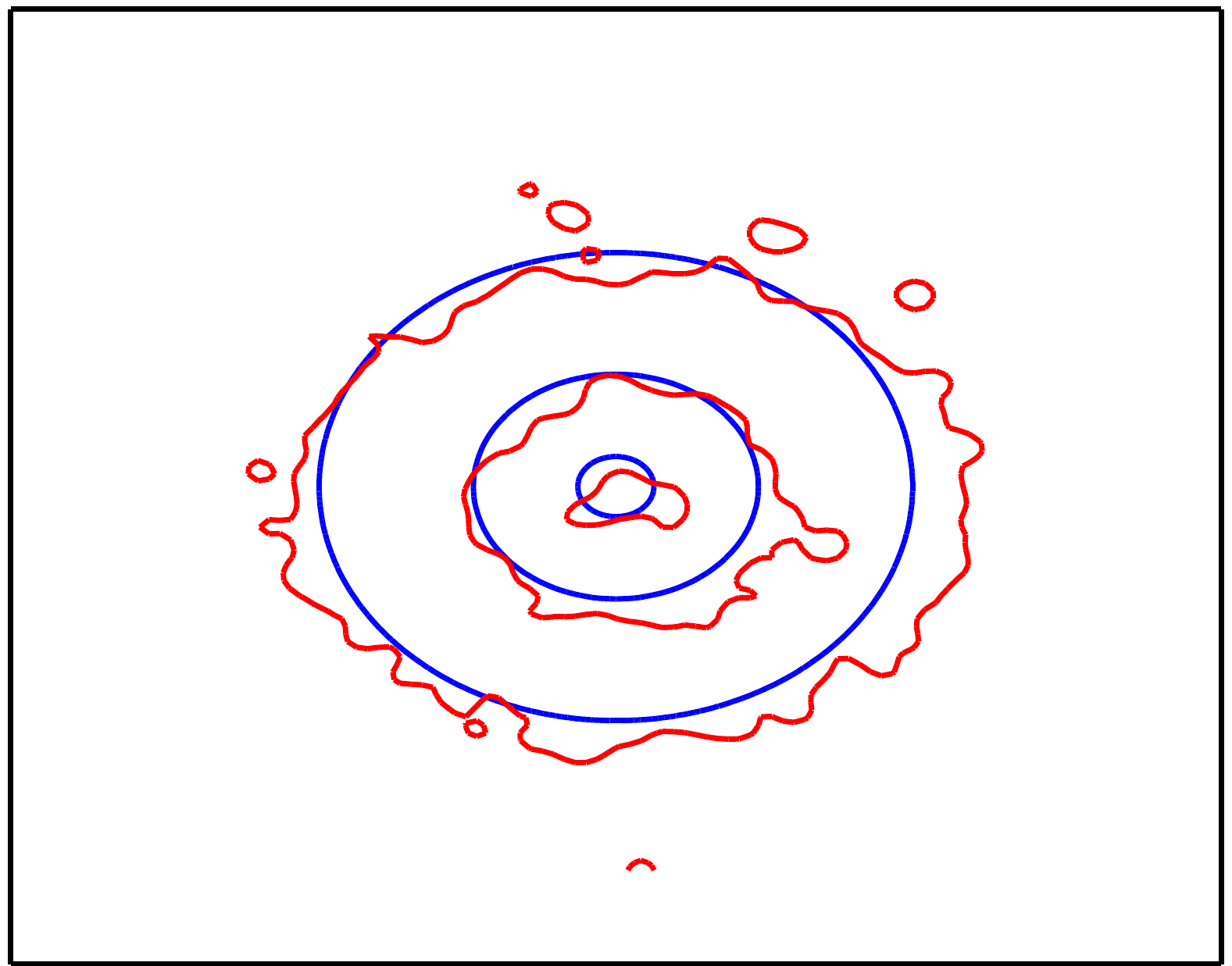}};

    \node (L41) at (0*\pos, 0*\posy)
          {\includegraphics[width=0.24\columnwidth]{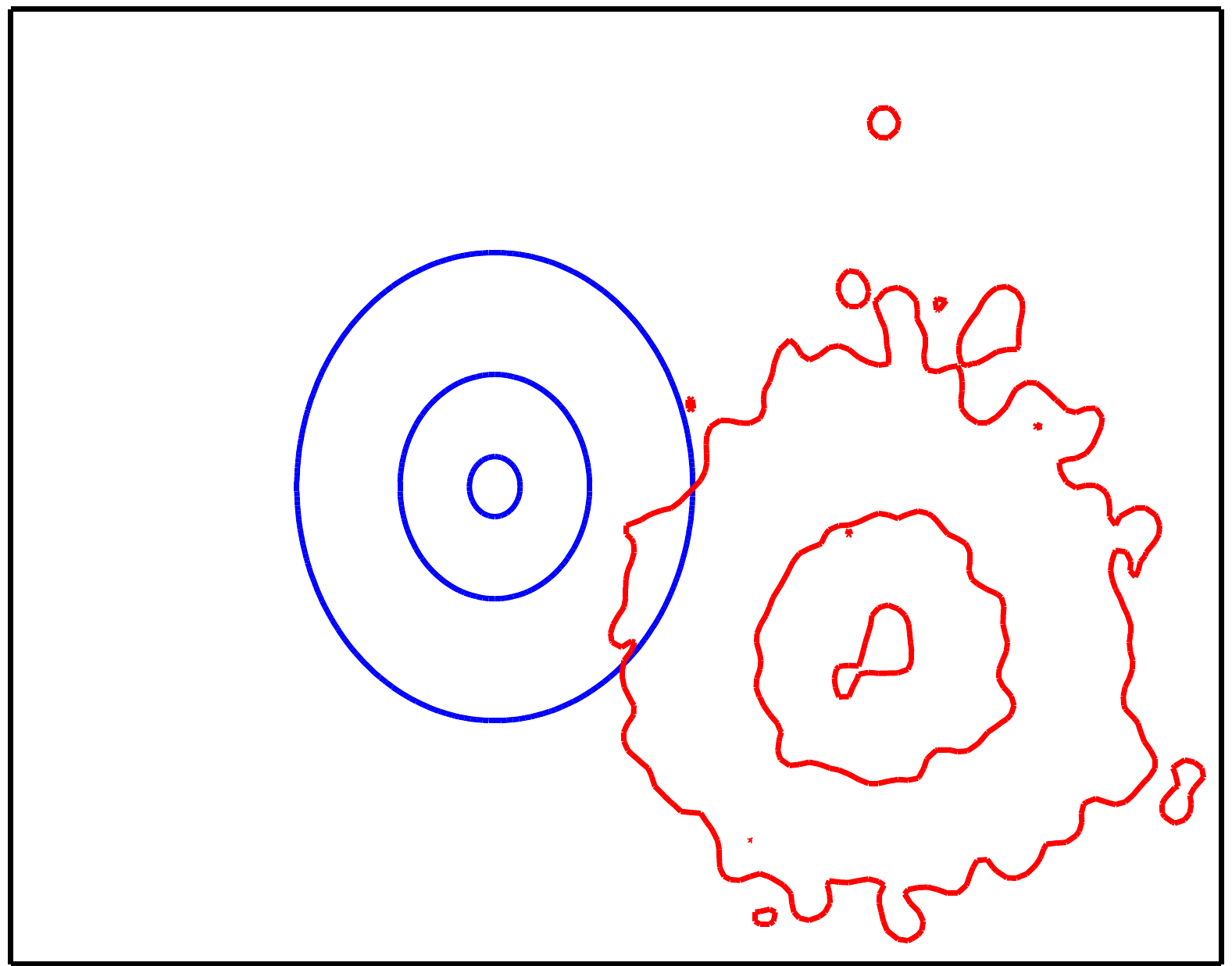}};
    \node (L42) at (1*\pos, 0*\posy)
         {\includegraphics[width=0.24\columnwidth]{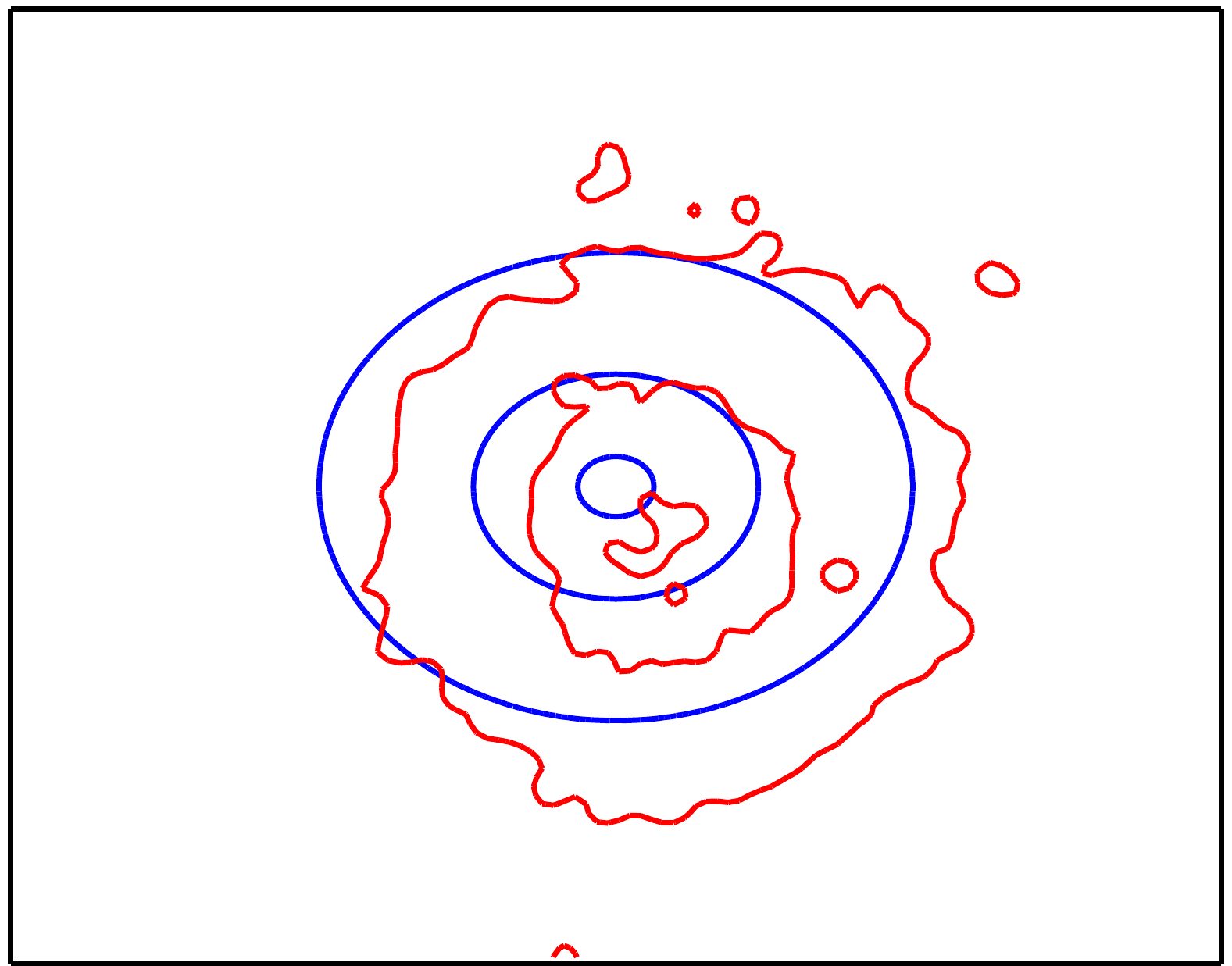}};
    \node (L43) at (2*\pos, 0*\posy)
         {\includegraphics[width=0.24\columnwidth]{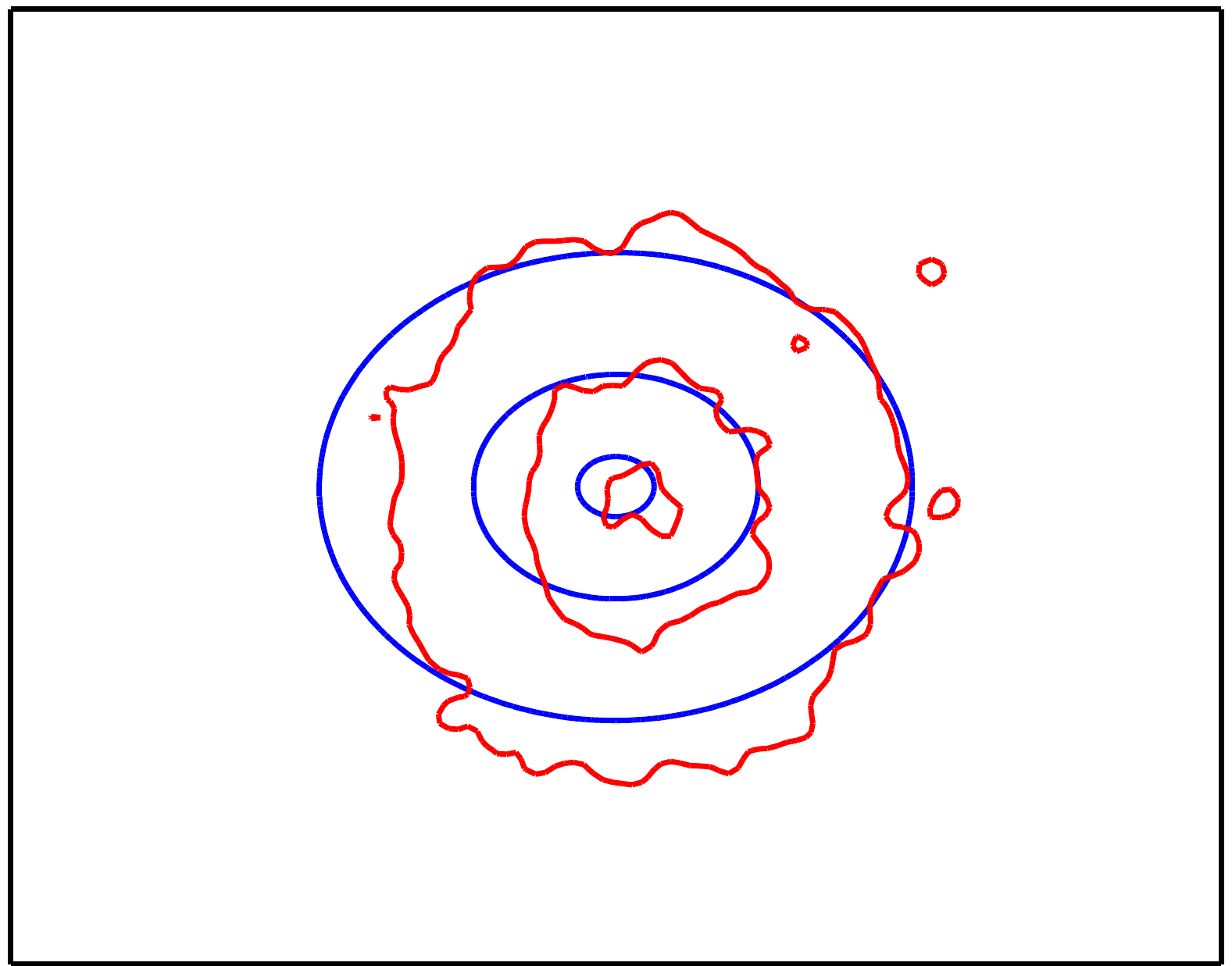}};

    \node (D1) at (0*\pos, 3*\posy)
          {\drawvarianceh{0.45}{1dmarginals_dataw1.dat}};
    \node (D2) at (1*\pos, 2*\posy)
          {\drawvarianceh{0.45}{1dmarginals_dataw3.dat}};
    \node (D3) at (2*\pos, 1*\posy)
          {\drawvarianceh{0.45}{1dmarginals_dataw18.dat}};
    \node (D4) at (3*\pos, 0*\posy)
          {\drawvarianceh{0.45}{1dmarginals_dataw26.dat}};

    \node [rotate=90] at (0*\pos-0.6*\pos,  2.75*\posy+0.29*\posy) {$v_{1}\downarrow$};
    \node [rotate=90] at (0*\pos-0.6*\pos,  1.75*\posy+0.29*\posy) {$v_{3}\downarrow$};
    \node [rotate=90] at (0*\pos-0.6*\pos,  0.75*\posy+0.29*\posy) {$v_{18}\downarrow$};
    \node [rotate=90] at (0*\pos-0.6*\pos, -0.25*\posy+0.29*\posy) {$v_{26}\downarrow$};

    \node at (0*\pos+0.0*\pos, 0*\posy-0.65*\posy) {$v_{1}\uparrow$};
    \node at (1*\pos+0.0*\pos, 0*\posy-0.65*\posy) {$v_{3}\uparrow$};
    \node at (2*\pos+0.0*\pos, 0*\posy-0.65*\posy) {$v_{18}\uparrow$};
    \node at (3*\pos+0.0*\pos, 0*\posy-0.65*\posy) {$v_{26}\uparrow$};

    \node at (2.5*\pos, 2.5*\posy)
          {\drawlegend{0.43}{2*\pos}{0.1*\posy}};
    \end{tikzpicture}
    \caption{One and two-dimensional marginals from the posterior
      (red) compared with marginals of the Gaussian approximation at
      the MAP point (blue).  The two-dimensional plots show contour
      lines of the two-dimensional marginals, where the three contours
      are selected to contain $5\%$, $50\%$, and $95\%$ of the
      density, respectively.  The marginals are computed with respect
      to the eigenvectors $v_1$, $v_3$, $v_{18}$ and $v_{26}$, which
      are plotted in Figure~\ref{fig:ordemodesh}a, b, d, and e,
      respectively. Kernel desity estimation is used to visualize the
      posterior pdf using the MCMC sample chain.
  \label{fig:2dmarginals_hesslike}}
  \end{figure}
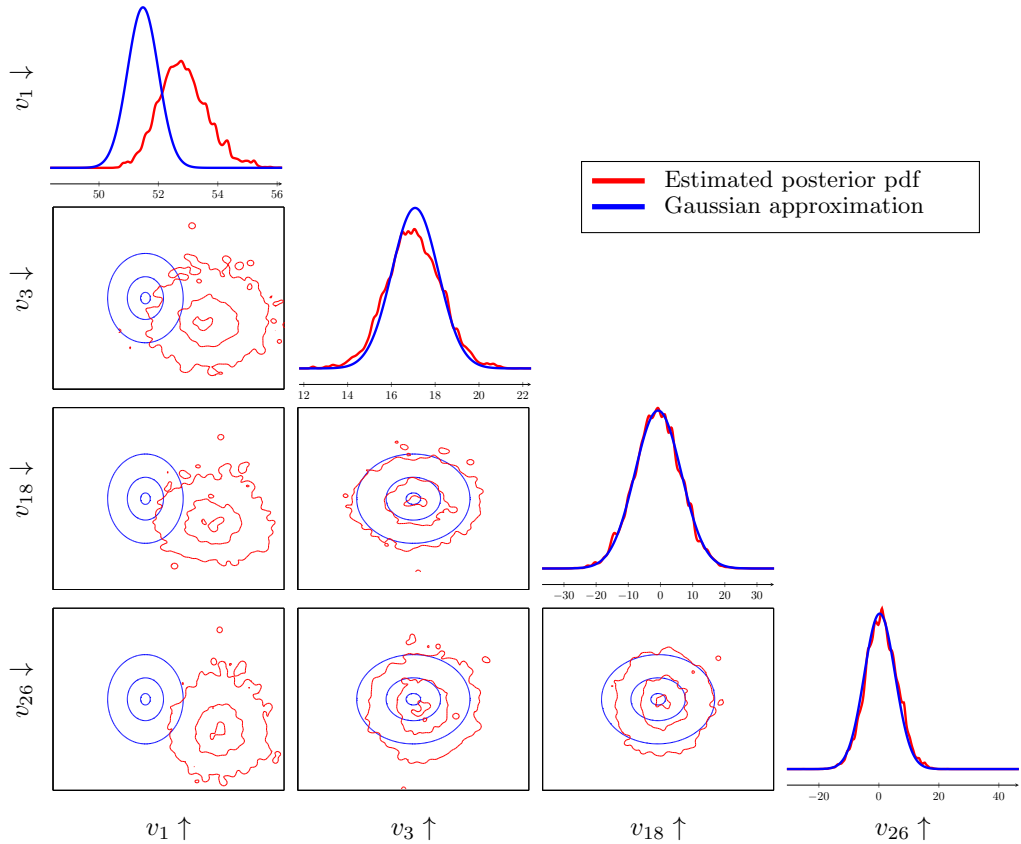
\end{center}

\section{Concluding remarks}
\label{sec:conclusions}

We have addressed the problem of constructing efficient MCMC methods
for exploring posterior distributions for uncertain parameter fields
in infinite-dimensional Bayesian inverse problems governed by
expensive forward models.
The stochastic Newton MCMC method presented in
\cite{MartinWilcoxBursteddeEtAl12} has been extended in several
ways. First, the method is recast in a form that is consistent with
the infinite-dimensional setting. In doing so, we have extended the
work in \cite{Bui-ThanhGhattasMartinEtAl13} to nonlinear inverse
problems. Second, the complexity of recomputing the Hessian at each
sample point was addressed by investigating a modified stochastic
Newton MCMC that reuses the Hessian evaluated at the MAP point.

The modified stochastic Newton MCMC method (with MAP-based Hessian)
proposed in this paper is compared with the original stochastic Newton
MCMC method (with dynamically changing Hessian) and with an
independence sampling method based on a Gaussian proposal at the MAP
point, for an ice sheet flow inverse problem governed by a nonlinear
Stokes equation. A performance comparison reveals that the proposed
stochastic Newton MCMC method with a MAP-based Hessian proposal leads
to the best convergence, both in terms of the number of samples as
well as in terms of the number of PDE solves.

We also presented visualizations and interpretations of the posterior
distribution in high dimensions. 
We showed point marginals of the posterior to provide intuition about
the statistical solution at particular points or regions of the
domain.  The point marginals confirm the dependence of the variance on
the availability of observations.
We classified the eigenvectors of the covariance of the Gaussian approximation of the posterior
at the MAP into groups depending on the extent to which they are influenced by
the observational data versus the prior. 
This classification can be used to identify and exploit directions in
parameter space in which the distribution is Gaussian (for directions
that are not informed by the data and hence are dominated by a
Gaussian prior) or non-Gaussian (for directions that are informed by
the data and hence the nonlinearity of the parameter-to-observable map
dominates).

\section*{Acknowledgements}
We would like to thank Youssef Marzouk for helpful discussions with
respect to the interpretation of the posterior distribution.

\bibliographystyle{siam}
\bibliography{ccgo}

\end{document}